\documentclass[pra,aps,superscriptaddress,twocolumn,floatfix,nofootinbib]{revtex4-2}

\usepackage{mathtools}
\usepackage[usenames,dvipsnames]{xcolor}
\usepackage{amsmath}
\usepackage{amssymb,mathrsfs}
\usepackage{graphicx}
\usepackage{dcolumn}
\usepackage{bm}
\usepackage{braket}
\usepackage{color}
\usepackage[colorlinks=true]{hyperref}
\usepackage{changes}

\def\br{{\bm r}}

\def\bu{{\bm u}}

\begin{document}

\title{Dissipation-enhanced collapse singularity of a nonlocal fluid of light in a hot atomic vapor}

\author{Pierre Azam}
\affiliation{Universit\'e C\^ote d'Azur, CNRS, Institut de Physique de Nice, Valbonne, France}

\author{Adrien Fusaro}
\affiliation{Laboratoire  Interdisciplinaire  Carnot  de  Bourgogne,  CNRS, Universit\'e  Bourgogne  Franche-Comt\'e,  Dijon,  France}
\affiliation{CEA, DAM, DIF, F-91297 Arpajon Cedex, France}

\author{Quentin Fontaine}
\affiliation{Laboratoire Kastler Brossel, Sorbonne Universit\'e, CNRS, ENS-PSL Research University, Coll\`ege de France, Paris, France}

\author{Josselin Garnier}
\affiliation{CMAP, Ecole Polytechnique, Institut Polytechnique de Paris, Palaiseau, France}

\author{Alberto Bramati}
\affiliation{Laboratoire Kastler Brossel, Sorbonne Universit\'e, CNRS, ENS-PSL Research University, Coll\`ege de France, Paris, France}

\author{Antonio Picozzi}
\affiliation{Laboratoire  Interdisciplinaire  Carnot  de  Bourgogne,  CNRS, Universit\'e  Bourgogne  Franche-Comt\'e,  Dijon,  France}

\author{Robin Kaiser}
\affiliation{Universit\'e C\^ote d'Azur, CNRS, Institut de Physique de Nice, Valbonne, France}

\author{Quentin Glorieux}
\affiliation{Laboratoire Kastler Brossel, Sorbonne Universit\'e, CNRS, ENS-PSL Research University, Coll\`ege de France, Paris, France}

\author{Tom Bienaim\'e}
\email[]{tom.bienaime@lkb.ens.fr}
\affiliation{Laboratoire Kastler Brossel, Sorbonne Universit\'e, CNRS, ENS-PSL Research University, Coll\`ege de France, Paris, France}

\date{\today}

\begin{abstract}
We study the out-of-equilibrium dynamics of a two-dimensional paraxial fluid of light using a near-resonant laser propagating through a hot atomic vapor.
We observe a double shock-collapse instability: a shock (gradient catastrophe) for the velocity, as well as an annular (ring-shaped) collapse singularity for the density.
We find experimental evidence that this instability results from the combined effect of the nonlocal photon-photon interaction and the linear photon losses.
The theoretical analysis based on the method of characteristics reveals the main result that dissipation (photon losses) is responsible for an unexpected enhancement of the collapse instability.
Detailed analytical modeling makes it possible to evaluate the nonlocality range of the interaction. The nonlocality is controlled by adjusting the atomic vapor temperature and is seen to increase dramatically when the atomic density becomes much larger than one atom per cubic wavelength. 
Interestingly, such a large range of the nonlocal photon-photon interaction has not been observed in an atomic vapor so far and its microscopic origin is currently unknown.
\end{abstract}

\maketitle

%%%%%%%%%%%%%%%%%%%%%%%%%%%%%%%%%%%%%%%%%%%%%%%%%%%%%%%%%%%%%%%%%%%%%%%
%                            Introduction                             %
%%%%%%%%%%%%%%%%%%%%%%%%%%%%%%%%%%%%%%%%%%%%%%%%%%%%%%%%%%%%%%%%%%%%%%%
\section{Introduction}

It has long been realized that light propagating in a nonlinear medium under the paraxial approximation can be interpreted in the context of quantum fluid dynamics \cite{Landau84,pomeau1993diffraction,frisch1992transition}. 
Such phenomena in nonlinear Kerr-like media that we recently call fluids of light have been investigated in photorefractive crystals \cite{Michel18, situ2020dynamics}, thermo-optic liquids \cite{Vocke15, Vocke16} and hot atomic vapors \cite{Santic18, Fontaine18, Fontaine20, piekarski2020short}.
In this work, we use a hot atomic vapor to study the impact of dissipation and the nonlocal character of the interactions in the shock-wave dynamics \cite{whitham}.
The theory of dispersive shock waves has been developed for a long time \cite{gurevich74}, 
after pioneering investigations in the fields of tidal waves \cite{benjamin54,peregrine66,johnson72} and collisionless plasma \cite{sagdeev79,zabusky65,taylor70}. 
It is only recently that dispersive shock waves have emerged as a general signature of singular 
fluid-type behavior \cite{biondini16,Onorato16,el16,miller16,Isoard19} in areas as different as 
Bose-Einstein condensates \cite{Dutton01,hoefer2006dispersive,chang2008formation}, shallow-water \cite{Trillo16}, oceanography \cite{smyth88}, plasma \cite{romagnani08}, 
viscous fluid conduits \cite{Maiden16} and several optical systems, 
e.g. photorefractive crystals \cite{Wan07,Gammal07,Ivanov20}, passive cavities \cite{Malaguti14}, and optical fibers \cite{Rothenberg89,Xu16,fatome14,garnier13,wetzel16,Trillo17,Xu17}.
An important ramification is the study of shock waves in the presence of a spatial nonlocal nonlinearity,
a feature studied experimentally in liquids with strong thermo-optic effects
\cite{Ghofraniha07,Conti09,Ghofraniha12,gentilini15,braidotti16,el16b,Karpov15,marcucci19,marcucci20}.
A nonlocal interaction means that the response of the nonlinearity at a particular point is not determined solely by the wave intensity at that point, but also depends on the wave intensity in its vicinity.
In this context, shock-wave formation from a random speckled beam revealed that a highly nonlocal nonlinearity leads to a double shock-collapse singularity \cite{Xu15,marcucci19}: 
while the gradient phase of the field (velocity) develops a shock, the intensity (density) develops a collapse instability on the edge of the ring-shaped beam, i.e., annular collapse.
On the other hand, we note that numerical simulations have shown the effect of nonlocal stabilization of nonlinear beams in a self-focusing atomic vapor \cite{skupin2007nonlocal}.
In these works, the dissipation has been considered as a negligible perturbative effect.

Here, we report the observation of the double shock-collapse instability in a fluid of light in an atomic vapor.
Our main result is to reveal a previously unrecognized and counterintuitive impact of the dissipation: The linear absorption of the optical field (\emph{i.e.} the fluid of light) is shown to be responsible for an enhancement of the collapse instability.
This unexpected result is enlightened by the theoretical analysis based on the method of characteristics for solving the hydrodynamic equations.
Our work then also contributes to the development of the concept of ``gain through losses"
in nonlinear optics, where specific frequency-distributed losses can be imaged into spectral gain \cite{tanemura04,perego18}.

An important aspect of our experimental work is to unveil a strong nonlocal regime for fluids of light propagating in hot atomic vapors. Contrary to previous studies on atomic vapors \cite{Tikhonenko96,Swartzlander92,abuzarli2021blast,bienaime2021controlled} where interactions were seen to be local, we tuned the atomic density up to $20$ atoms per cubic wavelength (by changing the temperature of the gas) which leads to an observable signature of nonlocality.
By comparing our experimental data to numerical simulations we give an estimate of the range of the nonlocal interactions in our system.
We propose a possible origin for such an unexpected large value of nonlocality reported in this work.

Our fluid of light is a continuous-wave laser beam which propagates in a nonlinear Kerr-like medium along the $z$ axis.
Under the paraxial approximation, the dynamics of the slowly varying amplitude $\psi$ of the laser electric field is commonly described by the nonlinear Schr\"odinger (NLS) equation \cite{Landau84}.
We consider the standard form of the nonlocal NLS equation: 
\begin{eqnarray}
	i \partial_z \psi &=& - \frac{\alpha}{2} \nabla^2 \psi - i \frac{\eta}{2} \psi \nonumber \\ 
                         & & + \gamma \psi  \int \mathrm d \br' U(\br -  \br') |\psi|^2(\br',z),
                         \label{NLSE}
\end{eqnarray}
where $\nabla$ is defined in the transverse plane $\br = (x , y)$ and the propagation axis $z$ plays the role of an effective time.
$\alpha=1/k_0$ is the dispersion (diffraction) parameter, where $k_0$ denotes the wave number of the laser and plays the role of an effective mass.
$\eta$ quantifies the strength of the linear absorption in the medium.
The interaction term includes an isotropic nonlocal response function $U(\br)={\mathcal{U}} ( |\br| )$ with the parameter $\gamma=  k_0 n_2$ that quantifies the strength of the interaction, $n_2$ being the Kerr nonlinear index.
We consider the repulsive photon-photon interactions (defocusing) regime $\gamma > 0$. 
We consider the examples of an exponential-shaped normalized response function $U(\br)= (2\pi \sigma^2)^{-1} \exp(-|\br|/\sigma)$, but similar results are obtained with a Gaussian-shaped response $U(\br)=(2\pi \sigma^2)^{-1} \exp(-|\br|^2/(2\sigma^2))$.
Note that, in addition to atomic vapors \cite{skupin2007nonlocal}, a nonlocal nonlinearity is found in several systems, e.g., dipolar Bose–Einstein condensates \cite{baranov08},  nematic liquid crystals \cite{peccianti12}, glasses \cite{Rotschild06}, liquids \cite{marcucci20}, or plasmas \cite{zakharov85}.

%%%%%%%%%%%%%%%%%%%%%%%%%%%%%%%%%%%%%%%%%%%%%%%%%%%%%%%%%%%%%%%%%%%%%%%
%                Experimental setup                                   %
%%%%%%%%%%%%%%%%%%%%%%%%%%%%%%%%%%%%%%%%%%%%%%%%%%%%%%%%%%%%%%%%%%%%%%%
%\section{Experimental setup}
\section{Experimental investigation}

\begin{figure}[]
\includegraphics[width=1\columnwidth]{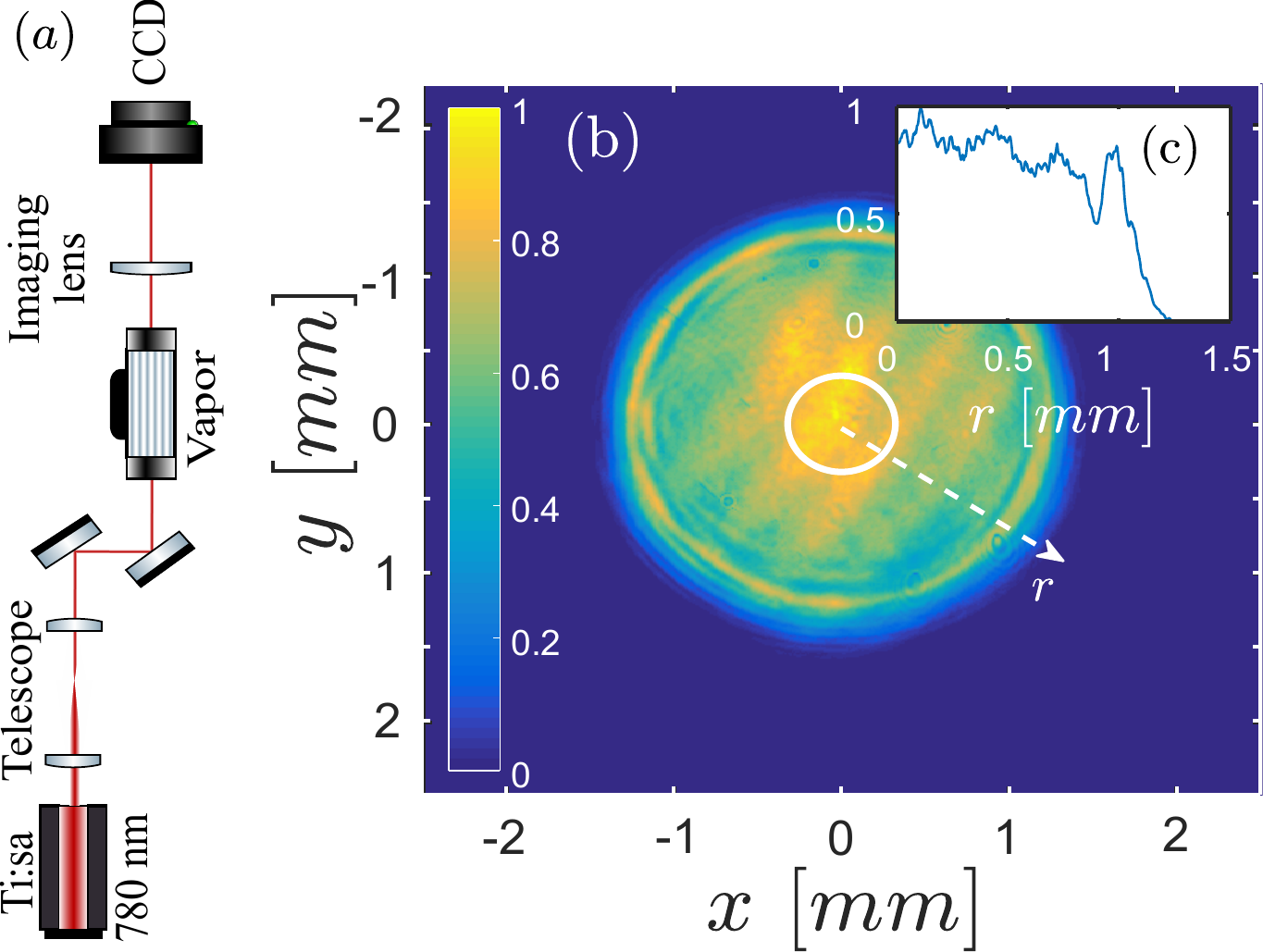}
\caption{(a) Simplified experimental setup. (b) Typical intensity profile obtained by imaging the output of the cell. The size of the beam at the entrance of the cell is illustrated by the white circle, whose radius equals the beam initial waist. Radial intensity profiles are extracted following the dashed line ($r-$axis). (c) Radial cut extracted from (b). }
\label{Fig1}
\end{figure}

We use an $L = 7 \, \text{cm}$ long cell filled with a gaseous natural isotopic mixture of $^{85}$Rb and $^{87}$Rb as nonlinear medium.
The fluid is created with a linearly polarized Ti:sapphire laser whose wavelength $\lambda$ is tuned near the $D_2$ resonance of $^{87}$Rb.
The laser detuning is adjusted from $-14\, \text{GHz}$ to $-2 \, \text{GHz}$ with respect to the $F=2 \rightarrow F'$ transition of $^{87}$Rb.
In this range, the detuning is large compared to the Doppler broadening ($\sim 250 \, \text{MHz}$) such that the Lorentzian shape of the line dominates.
The atomic vapor density is controlled by adjusting and stabilizing the temperature of the cell from $100 \, ^{\circ}\text{C}$ to $160 \, ^{\circ}\text{C}$, which corresponds to a range of one $^{87}$Rb atom per $\lambda^3$ to $20$ atoms per $\lambda^3$ respectively.
With these two parameters (laser detuning and temperature), we can adjust the nonlinear Kerr index $n_2$ and thus the effective photon-photon interaction $\gamma$. 
To calibrate the value of $n_2$, we measure the far field intensity of a collimated Gaussian beam with initial waist of $0.7 \, \text{mm}$ and peak intensity $I=6.5 \, 10^5 \, \text{W}\, \text{m}^{-2}$.
Due to self-phase modulation, this configuration generates concentric rings and provides a direct measurement of the nonlinear phase $\phi_{\text{NL}} = k_0 n_2 I L$ accumulated by the beam along its propagation.
In this work, we measure values of $n_2$ from $1 \times 10^{-11} \, \text{m}^2 \, \text{W}^{-1}$ to $3.8 \times 10^{-10}  \, \text{m}^2 \, \text{W}^{-1}$ \cite{Zhang15}. 

The experimental setup is depicted in Fig. \ref{Fig1}(a).
We set the initial dimension of the fluid of light by demagnifying the beam down to a waist of $w = 0.32 \, \text{mm}$.
The waist is located at $z=0$ and the Rayleigh length of the initial beam $z_{\text{R}} = \pi w^2 / \lambda \simeq 40  \, \text{cm}$ exceeds the length of the cell.
The total power can be adjusted from $0$ to $1 \, \text{W}$, leading to a peak intensity between $0$ and  $I = 6.2 \times 10^6 \, \text{W}\, \text{m}^{-2}$.
In  Fig. \ref{Fig1}(b), we present a typical image of the cell output intensity compared to the input beam size.
A clear difference between the initial and the final dimensions of the fluid is visible, due to the repulsive photon-photon interaction.
Fig. \ref{Fig1}(c) shows a typical radial cut of the intensity profile that we use to study the dynamics of the fluid.\\

\begin{figure}[]
\centerline{{\includegraphics[width=0.45\textwidth]{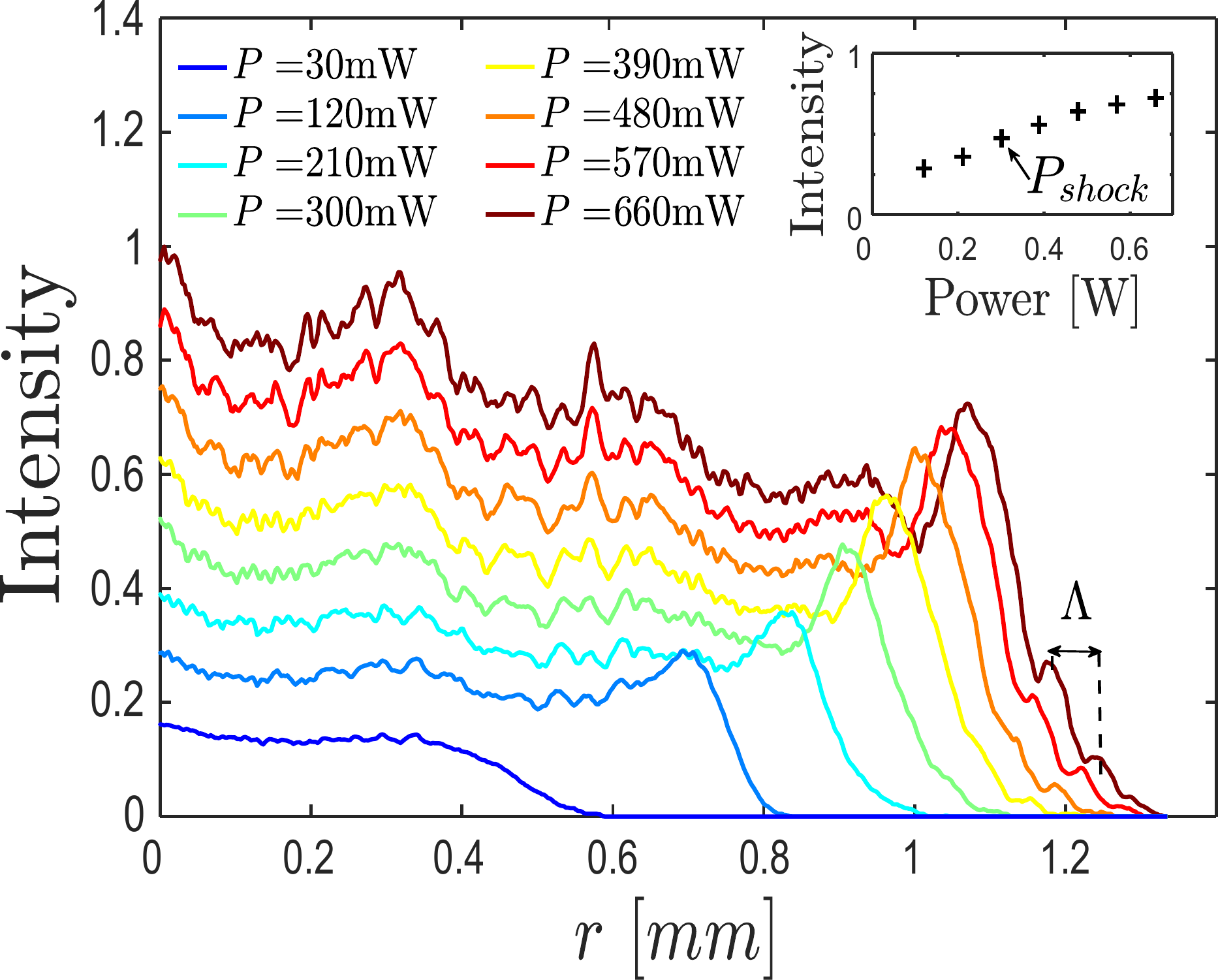}}}
	\caption{Output radial  intensity profile for different input powers. The cell is heated to $150 \, ^\circ \text{C}$, the laser detuning is $\Delta = - 5.5 \, \text{GHz}$ and the transmission is $\mathcal{T}=40\%$. By looking at the different profiles we can identify the critical power $P_{\text{shock}}=300 \, \text{mW}$ beyond which the front profile is characterized by  the formation of the rapidly oscillating density, which regularizes the shock  with a spatial period given by the healing length $\Lambda$. Inset: Maximum intensity of the collapse peak as a function of the beam power. It hints at a faster-than-linear growth (resp. slower) below $P_{\text{shock}}$  (resp. beyond) in agreement with our theoretical analysis.}
\label{Fig2}
\end{figure}

%%%%%%%%%%%%%%%%%%%%%%%%%%%%%%%%%%%%%%%%%%%%%%%%%%%%%%%%%%%%%%%%%%%%%%%
%                Experimental results                                   %
%%%%%%%%%%%%%%%%%%%%%%%%%%%%%%%%%%%%%%%%%%%%%%%%%%%%%%%%%%%%%%%%%%%%%%%
%\section{Experimental results}

The experimental protocol consists of fixing the laser transmission by adjusting the laser detuning $\Delta$ after the cell has been heated to a given temperature $T$.
Then we explore the dynamics of the fluid of light by varying the beam power.
The radial intensity profiles are shown in Fig. \ref{Fig2}. 
Here we show data for a cell temperature of $150 \, ^\circ \text{C}$ and a transmission of $40 \%$. 
From each dataset (temperature and transmission) we can identify the development of a shock-wave.
For early effective times, we observe a clear self-steepening of the shock-front, as expected for a conventional shock-wave \cite{Trillo17}. 
Note that, in our experiment the shock-wave builds even in the absence of a background fluid to sustain it \cite{Xu16}.
This is confirmed for late times where the wave breaking is regularized by the dispersion effects and the formation of the characteristic rapidly oscillating density front.
%\deleted{, which defines the shock point with power $P_{\rm shock}$}. 
While the appearance of a shock is theoretically associated with a gradient catastrophe of the velocity, since our experimental observable is the density we define 
the critical power $P_{\rm shock}$ as the beam power beyond which a dispersive, oscillating structure emerges at the front of the density profile.
In Fig.~\ref{Fig2} we can see that $P_{\rm shock}=0.3$W,
%\deleted{These two criteria are direct consequences on the density observable of the velocity gradient catastrophe. We also note that,}
and we can also note that, as expected, the spatial period of the oscillating front is typically given by the healing length $\Lambda$, which is the spatial length scale for which linear and nonlinear effects are of the same order. In the case of the NLS Eq.(\ref{NLSE}) we have $\Lambda = 1/\sqrt{2 k_0 \gamma I_f}$,  where we have considered the local value of the intensity $I_f$ at the shock front. We obtain  $\Lambda \simeq 50 \, \mu$m in the case of Fig.~\ref{Fig2}.

\begin{figure}[]
\centerline{{\includegraphics[width=0.45\textwidth]{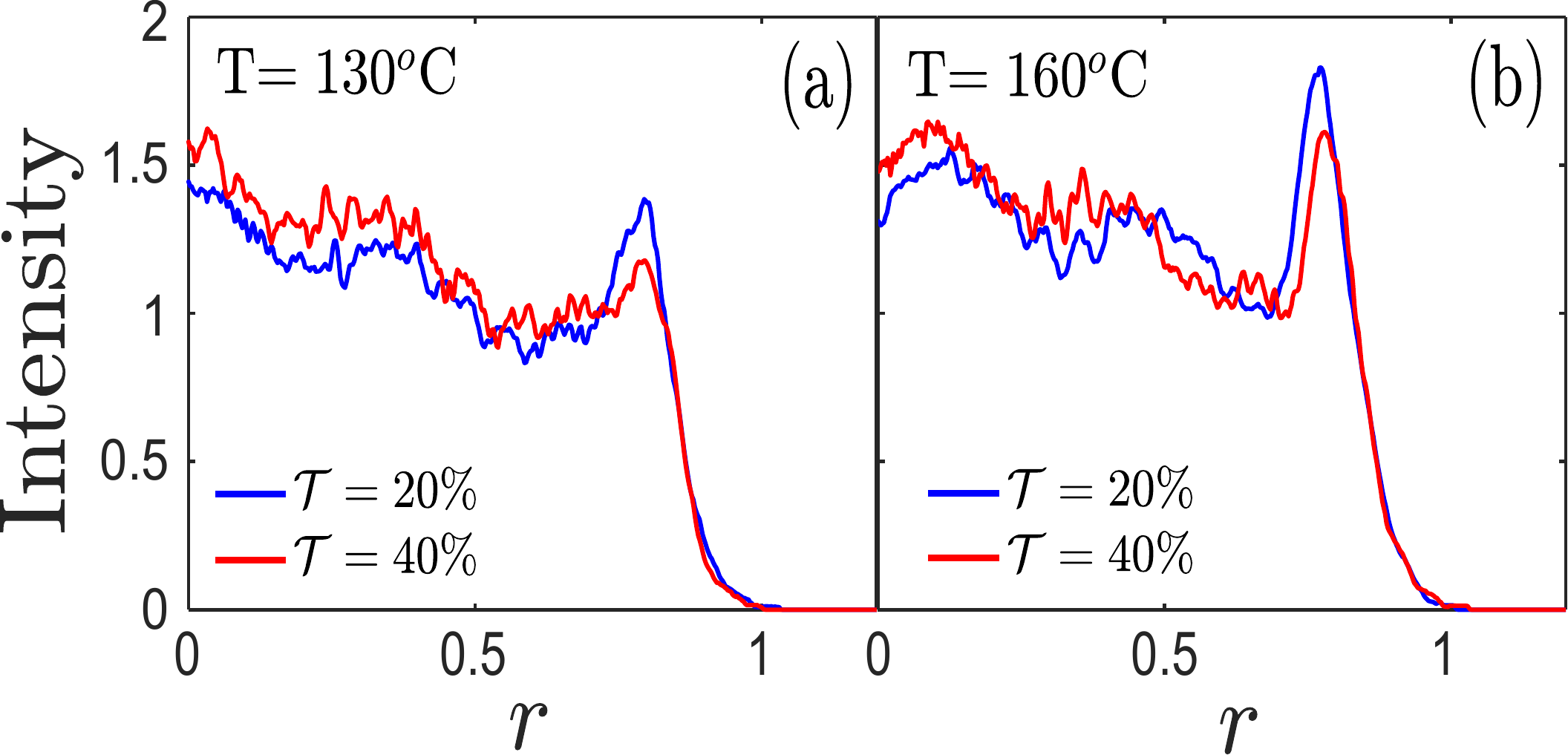}}}
	\caption{Intensity profiles at the critical power showing that the collapse instability is enhanced by increasing either the dissipation, or the nonlocality (i.e. temperature).
After fixing the nonlocality thanks to the vapor temperature, transmissions are chosen using the frequency detuning. The power at which the shock occurs is  identified as explained through Fig. \ref{Fig2}. (a) For  $T=130 \, ^\circ \text{C}$, the detunings at $\mathcal{T}=40\%$ and $20\%$ are respectively $-3.3$ and $-2.1 \, \text{GHz}$. (b) For  $T=160 \, ^\circ \text{C}$, the detunings at $\mathcal{T}=40\%$ and $20\%$ are respectively $-8.2$ and $-5.4 \, \text{GHz}$. Spatial and intensity profiles have been renormalized to improve the comparison of the collapse peak amplitude.}
\label{Fig3}
\end{figure}

At the critical power $P_{\rm shock}$, the beam exhibits a pronounced peak intensity on the ring-shaped shock-front.
This observation is similar to that reported in liquids with strong thermo-optic 
effects  \cite{braidotti16,marcucci19}.
It is also apparently similar to the collective incoherent 
shock dynamics of speckle beams where the annular instability on the shock front was recognized as a collapse singularity \cite{Xu15}.
Such an instability is termed `annular' collapse because it does not occur as usual at a single spatial point, but on the circular edge of the ring-shaped beam.
The main result of our experiments is that such 
an annular peak intensity is enhanced (i) at high temperature (high nonlocality) and (ii) at low transmission 
(high dissipation), see Fig.~\ref{Fig3}.
While the role of the vapor temperature can be understood by an increase of the nonlocal range of the interactions, the impact of dissipation has not been discussed in previous studies on shock waves.
Here we stress the crucial impact of the dissipation in the development of the collapse instability since the annular collapse would be almost invisible in our experiments if the impact of the dissipation could be neglected.

%%%%%%%%%%%%%%%%%%%%%%%%%%%%%%%%%%%%%%%%%%%%%%%%%%%%%%%%%%%%%%%%%%%%%%%
%                            Results and Simulations                  %
%%%%%%%%%%%%%%%%%%%%%%%%%%%%%%%%%%%%%%%%%%%%%%%%%%%%%%%%%%%%%%%%%%%%%%%
\section{Theory, simulations and discussions}

In this section, we explain why the nonlocal nonlinearity combined with dissipation has a dramatic impact on the fluid dynamics.
Using the method of characteristics and numerical simulations, we demonstrate that the presence of dissipation changes the nature of the singularity and leads to the formation of a double shock collapse singularity.

Starting from the NLS Eq.(\ref{NLSE}), we make the change of variable $\Psi(\br,z) = \psi(\br,z) \exp( \eta z/2)$ to get 
\begin{align}
    i \partial_z \Psi =& -\frac{\alpha}{2} \nabla^2 \Psi \nonumber \\
    &+ \gamma e^{-\eta z} \Psi  \int U(\br-\br' )   |\Psi|^2 (\br',z) {\mathrm d}\br'.
    \label{eq:nls_Psi}  
\end{align}
Following the Madelung transformation, the wave amplitude factors as $\Psi(\br,z)=\sqrt{\rho(\br,z)} \exp\big(i\phi(\br,z)\big)$, and the NLSE (\ref{eq:nls_Psi}) takes the form:
\begin{eqnarray}
\partial_z \rho + \alpha \nabla\cdot(\rho \nabla \phi) &=& 0 , 
\label{eq:rho_0}\\
\partial_z \phi +\frac{\alpha}{2} (\nabla \phi)^2 + e^{- \eta z}V &=& \frac{\alpha}{2 \sqrt{\rho}} \nabla^2 \sqrt{\rho},
\label{eq:phi_0}
\end{eqnarray}
with the potential
\begin{eqnarray}
V(\br,z)= \gamma \int U(\br - \br') \rho(\br',z) {\mathrm d}\br'.
\label{eq:V}
\end{eqnarray}
%The last term in (\ref{eq:phi_0}) refers to the analogue of the Bohm quantum potential.
Defining an analogue of the `velocity' $\bu=\nabla \phi$, we get the following  dispersive hydrodynamic-like equations
\begin{eqnarray}
&&\partial_z \rho +  \alpha \nabla \cdot \big(\rho \, \bu \big) = 0,
\label{eq:Nvec0}\\
&&\partial_z \bu + \alpha (\bu \cdot \nabla) \bu + e^{-\eta z}\nabla V = \frac{\alpha}{2} \nabla \big( \frac{1}{\sqrt{\rho}} \nabla^2 \sqrt{\rho} \big).
\label{eq:uvec0}
\end{eqnarray}
In the experiment, the initial Gaussian beam is radially symmetric with $\bu(\br,z=0) ={\bf 0}$. 
In cylindrical coordinates, the momentum is radially outgoing $\bu = u(r,z) \br/r$ with $r=|\br|$, while  $\rho = \rho(r,z)$ is independent of the polar angle $\theta$.
Indeed, such a  radial symmetry is verified by the initial condition and the dispersive hydrodynamic equations preserve this property during the propagation in $z$.
The hydrodynamic equations (\ref{eq:Nvec0}-\ref{eq:uvec0}) can thus be reduced to the effective 1D-radial system:
\begin{align}
&\partial_z \hat{\rho} +\alpha \partial_r( \hat{\rho}  u ) = 0,
\label{eq:N1dr}\\
&\partial_z u +  \alpha u \partial_r u +
e^{-\eta z} \partial_r V =
\frac{\alpha}{2} \partial_r \big( \frac{1}{\sqrt{r} \sqrt{\hat{\rho}}} \partial_r r \partial_r
\frac{ \sqrt{\hat{\rho}}}{\sqrt{r}} \big),
\label{eq:u1dr}
\end{align}
where $\hat{\rho}(r,z) = r \rho(r,z)$, and the potential is given by
\begin{eqnarray}
V (r,z ) = \gamma \int_0^\infty \tilde{U}(r,r')  \rho(r',z) r'{\mathrm d} r',
\label{Vr}
\end{eqnarray}
with $\tilde{U}(r,r') = \int_0^{2\pi} {\mathcal{U}} \big(\sqrt{r^2+{r'}^2-2rr' \cos \theta} \big) {\mathrm d} \theta$.

Considering the strongly nonlinear regime inherent to the development of shock-waves, the impact of dispersion effects in the rhs of (\ref{eq:u1dr}) can be neglected before the occurrence of the singularity.
In this non-dispersive regime, Eqs.(\ref{eq:N1dr}-\ref{eq:u1dr}) take the form of hydrodynamic-like equations accounting for the nonlocality and the dissipation:
\begin{align}
&\partial_z \hat{\rho} +\alpha \partial_r( \hat{\rho}  u ) = 0,
\label{eq:N1dr_hyd}\\
&\partial_z u +  \alpha u \partial_r u + e^{-\eta z}\partial_r V = 0.
\label{eq:u1dr_hyd}
\end{align}

Before pursuing, it proves convenient to briefly recall the usual local limit without dissipation.
In this regime, Eq.~(\ref{eq:u1dr_hyd}) reduces to $\partial_z u + \alpha u \partial_r u + \gamma \partial_r \rho = 0$.
Starting from $u(r,z=0)=0$, the velocity $u(r,z)$ is first driven by the last  term in (\ref{eq:u1dr_hyd}).
Then the dynamics is dominated by the Hopf (or inviscid Burgers) second term in  (\ref{eq:u1dr_hyd}), which leads to the gradient catastrophe of the velocity $u(r,z)$.
When combined to (\ref{eq:N1dr_hyd}), the system then develops a radial symmetric  annular shock that is characterized by a divergence of $\partial_{r} u$ and $\partial_{r} \rho$, while $\rho$ remains finite and does not blow up \cite{whitham}.
We recall that the system does not develop, strictly speaking, a singularity at finite propagation distance, because the neglected dispersive terms (rhs in (\ref{eq:u1dr})) become important nearby the shock distance (the propagation distance that corresponds to the onset of the dispersive, oscillating structure at the front of the density profile). The dispersive effects regularize the singularity \cite{Trillo17}, i.e., the system does not develop a shock in the Rankine-Hugoniot sense \cite{whitham}.

\begin{figure}[]
\centerline{{\includegraphics[width=0.5\textwidth]{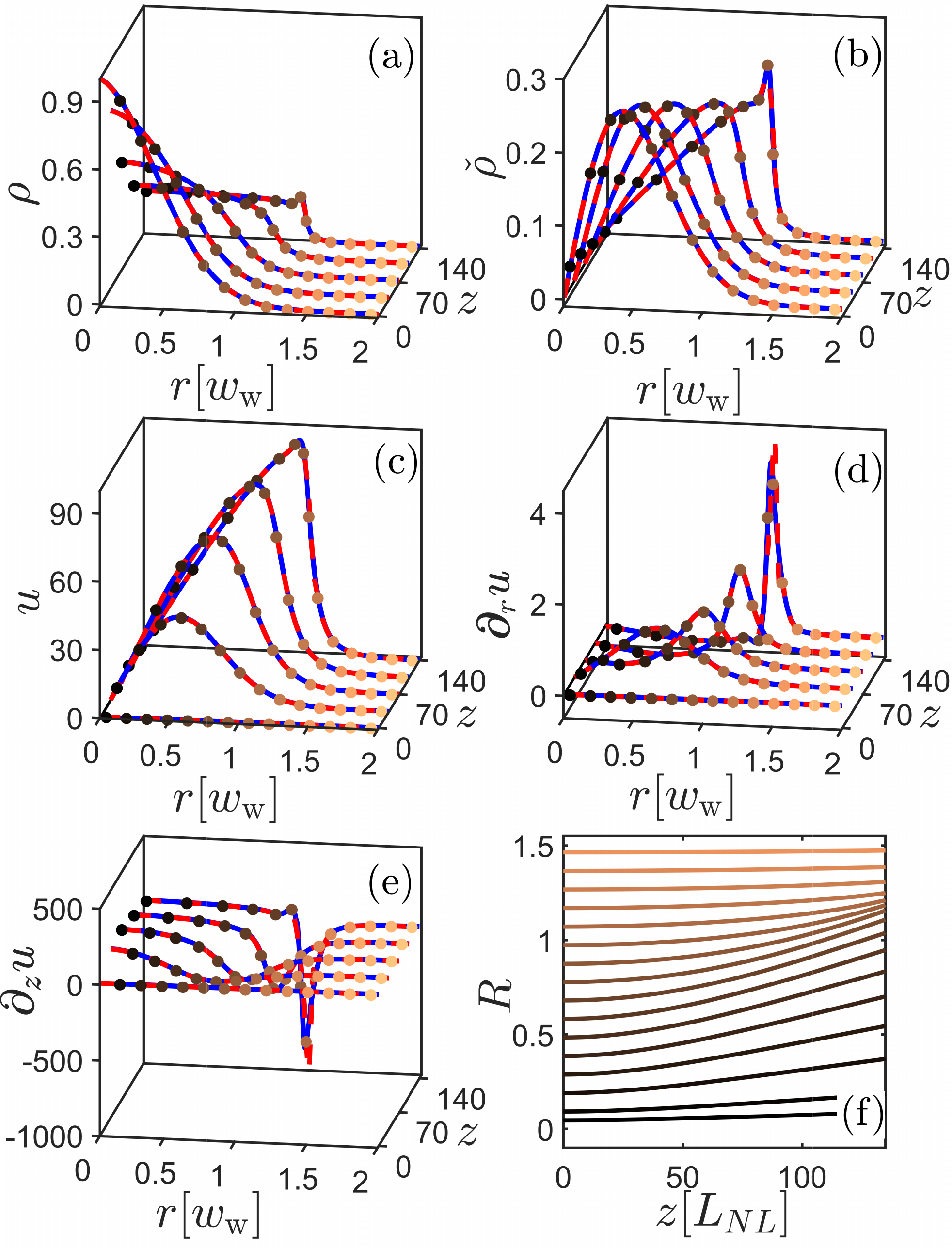}}}
	\caption{\baselineskip 10pt
(a-e) Simulations of the NLS Eq.(\ref{NLSE}) (dashed red), of the reduced hydrodynamic model (\ref{eq:N1dr}-\ref{eq:u1dr}) (solid blue), and of the system of characteristic equations (\ref{caracgen1}-\ref{caracgen5}) (filled circles), for $z = 0, 30, 65, 100, 135$ in units of $z_{NL}=1/(\gamma I)$.
Evolution of the density $\rho(r,z)$ (a), the radial density $\hat{\rho}(r,z)$ (or ${\check \rho}(z)$) (b), the velocity $u(r,z)$ (or $w(z)$) (c), the gradient of the velocity $\partial_r u(r,z)$ (or $\xi(z)$) (d) and the ‘time’ $z-$derivative of the velocity $\partial_z u(r,z)$ (or $\tau(z)$) (e), for different characteristics $R(z)$ (f).
The colors of the characteristics in (f) correspond to the colors of the filled circles in (a)-(e). 
The characteristics tend to approach each other nearby the shock point. 
Correspondingly, the velocity $u(r,z)$ (c) exhibits a self-steepening process followed by a shock singularity (i.e., a collapse for the  gradient $\xi(r,z)$) (d), which in turn induces a collapse singularity for the density ${\hat \rho}(r,z)$ (b).
Parameters are $\sigma=50 \, \mu$m and $\mathcal{T}=60$\% after propagation through the cell.
See the Movie in \cite{supplemental} for a visualization of the whole dynamics.
}
\label{Fig4}
\end{figure}

The scenario discussed here above also holds in the weakly nonlocal regime $\sigma \ll \Lambda$.
On the other hand, in the regime $\sigma > \Lambda$, the nonlocality changes the nature of the singularity.
The main observation is that, at variance with $\partial_{r} \rho$ that diverges, the term $\partial_r V$ in (\ref{eq:u1dr_hyd}) exhibits a regular behavior around the shock position.
This regularization of $\partial_r V$ has a dramatic effect: in the absence of nonlocality ($\sigma=0$), the term $\gamma \partial_r \rho$ was preventing from the growth of the amplitude of $\rho$, but without such a term, the amplitude of $\rho$ does grow.

This behavior is revealed by applying the method of characteristics \cite{evans} to the hyperbolic system  Eqs.(\ref{eq:N1dr_hyd}-\ref{eq:u1dr_hyd}). 
We define $w(z)=u(R(z),z)$, $\tau(z)= \partial_z u\big(R(z),z\big)$, $\xi(z)= \partial_r u\big(R(z),z\big)$, $\check{\rho}(z) =  \hat{\rho}\big(R(z),z\big)$ by following a specific characteristic $R(z)$ with $R(0)=r_0$.
These functions can be shown to satisfy the following system of ordinary differential equations (ODEs):
\begin{eqnarray}
&&\frac{dR}{dz} = \alpha w(z) ,   
\label{caracgen1} \\
&&\frac{dw}{dz} =  \tau(z)+  \alpha w(z)\xi(z)  ,  
\label{caracgen2} \\
&&\frac{d\tau}{dz} = -\partial^2_{zr} (e^{-\eta z}V(R(z),z)) -  \alpha \xi(z)\tau(z), 
\label{caracgen3} \\
&&\frac{d\xi}{dz} = -\partial^2_{r} (e^{-\eta z}V(R(z),z))  - \alpha \xi^2(z) , 
\label{caracgen4} \\
&&\frac{d\check{\rho}}{dz} =  - \alpha \xi (z)\check{\rho} (z).
\label{caracgen5}
\end{eqnarray}
From the formal point of view, these equations are analogous to those derived in the framework of the long-range Vlasov formalism in the strong turbulence regime \cite{Xu15,xu16incoherent,xu18incoherent}, or to describe the emergence of phase coherence from a random speckled beam  \cite{fusaro17}. Note that the impact of dissipation has not been discussed in these previous works.

The development of the double shock collapse singularity is driven by the velocity gradient in Eq.(\ref{caracgen4}).
In the first term of (\ref{caracgen4}), the amplitude of $\partial_r^2 V $ can be bounded by the maximal amplitude of $\hat{\rho}$ divided by $\sigma^2$.
On the other hand, the second term in (\ref{caracgen4}) describes a divergence of the form $|\xi| \sim 1/(z_\infty-z)$ close to the shock point $z_\infty$.
Thus, there exists some propagation length beyond which the term $-\alpha \xi^2$ dominates $\partial_r^2 V(r,z)$, and $\xi(z)$ blows up in finite effective time $z$.
Remarking furthermore that $\check{\rho}(z) = \hat{\rho}(r_0,z_0) \exp \big( - \alpha \int_{z_0}^z \xi(s) {\mathrm d} s \big)$ from (\ref{caracgen5}), we obtain the singular behaviors of $\xi(z)$ and $\check{\rho}(z)$ just before $z=z_\infty$:
\begin{eqnarray}
&& \partial_r u\big(R(z),z\big) \simeq -1/[\alpha(z_\infty-z)],\\
&& \hat{\rho}\big(R(z),z\big) \simeq \hat{\rho}(r_0,z_0) (z_\infty-z_0)/ (z_\infty-z).
\end{eqnarray}

Note the quantitative agreement obtained in Fig.~\ref{Fig4}(a)-(e) between the simulations of the NLS Eq.(\ref{NLSE}), the hydrodynamic Eq.(\ref{eq:N1dr}-\ref{eq:u1dr}), and the characteristic Eqs.(\ref{caracgen1}-\ref{caracgen5}), without using adjustable parameters. Note that we considered a Gaussian-shaped response function.
We remark that the singularity is not regularized by the dissipation (as for conventional diffusive shocks \cite{whitham}), but by the dispersion effects neglected in Eqs.(\ref{eq:N1dr}-\ref{eq:u1dr}), which leads to the formation of the characteristic rapidly oscillating density front observed in the experiments, see Fig.~\ref{Fig2}.

\begin{figure}[]
\centerline{{\includegraphics[width=0.5\textwidth]{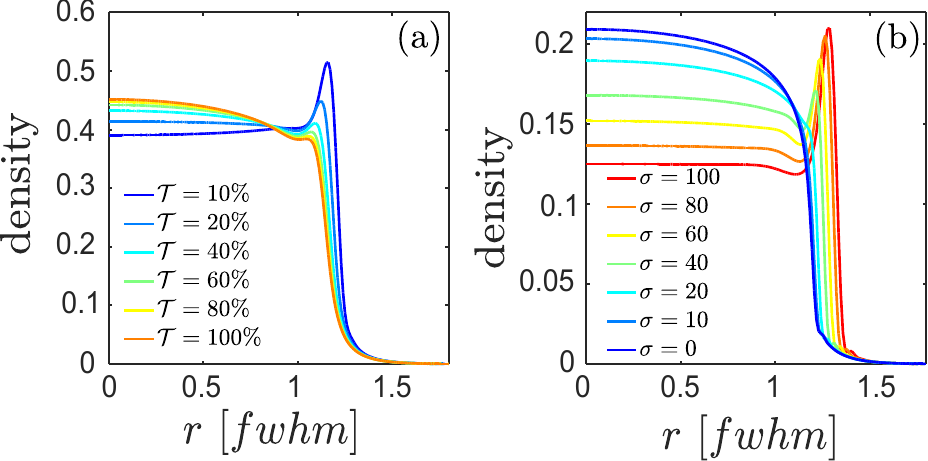}}}
	\caption{\baselineskip 10pt
(a) Density profile $\rho(r)$ for different values of the transmissions $\mathcal{T}$ (i.e., different $\eta$), but for the same effective nonlinear propagation length $z_{eff}=(1-\exp(-\eta z))/\eta=120 \, z_{NL}$ ($\sigma=70 \, \mu$m).
(b) Density profile $\rho(r)$ for different values of nonlocality $\sigma$ (in microns) nearby the shock point $z_\infty$ ($\mathcal{T}=60$\%).
Note that, to improve the visualization, all plots in (a) have been normalized to the power for $\mathcal{T}=100$\% (orange line).
}
\label{Fig5}
\end{figure}

This analysis unveils the unexpected role of the dissipation. 
By quenching the first term in Eq.~\eqref{caracgen4}, dissipation  $\eta$ favors the growth of $|\xi|$ described by the second term, which in turn favors the growth of $\check{\rho}(z)$ before the singularity.
This effect is illustrated in Fig.~\ref{Fig5}(a) where the dissipation is seen to strengthen the development of the collapse singularity at fixed nonlocality.
Our experimental data, shown in Fig.~\ref{Fig3}, confirm this prediction since the amplitude of the collapse instability is enhanced by slightly decreasing the transmission from 40\% to 20\%.
Similarly, our results show that the nonlocality favors the development of the shock-collapse singularity by quenching the bounds of $\partial_r^2 V(r,z)$ in (\ref{caracgen4}).
This is illustrated in Fig.~\ref{Fig5}(b) that reports the density profiles nearby the shock point for different mounts of nonlocality.

Using this model, we estimate the amount of nonlocality required to describe quantitatively our experimental results.
We find a nonlocality of $\sigma \simeq 70 \, \mu \text{m}$ for a temperature of $160 \, ^\circ$C.
Such a large nonlocality coefficient is unexpected since, in hot atomic vapors, nonlocality is usually attributed to the ballistic transport of fast-moving excited atoms, which results in a nonlocal length scale of the order of $\sigma \simeq 7.5 \, \mu \text{m}$ for $160 \, ^\circ$C \cite{tam1977long,suter1993stabilization,skupin2007nonlocal}.
The larger nonlocal length scale needed to explain our observation requires to go beyond the simple two-level model used in \cite{skupin2007nonlocal} and includes more complex atom-light interaction, including for instance possible optical pumping \cite{Maucher_2016} with long-range memory, the diffusive motion of atoms \cite{suter1993stabilization}, collisional broadening \cite{VanKampen97} and other collective optical response mechanisms.

%%%%%%%%%%%%%%%%%%%%%%%%%%%%%%%%%%%%%%%%%%%%%%%%%%%%%%%%%%%%%%%%%%%%%%%
%                    Conclusion                                       %
%%%%%%%%%%%%%%%%%%%%%%%%%%%%%%%%%%%%%%%%%%%%%%%%%%%%%%%%%%%%%%%%%%%%%%%

\section{Conclusion}

In summary, we have reported the observation of a  shock-collapse instability for the evolution of velocity and the intensity of an optical field propagating  in an atomic vapor.
We have shown experimentally and theoretically that the linear dissipation combined to the nonlocal nonlinearity is responsible for a significant enhancement of the collapse instability.
Furthermore, the theoretical analysis can be applied to a purely local nonlinear interaction, which remarkably reveals that the shock formation (density gradient collapse) is associated with a density collapse only if there is dissipation.
Our experimental platform enables the study of the crossover from a local nonlinear interaction to a nonlocal one by tuning the temperature of the vapor cell.
We observed a nonlocal interaction which is an order of magnitude larger than previously reported for ballistic transport. 
These results open the way to tunable nonlocal physics with fluids of light and require multi-level atoms modelization to identify the microscopic mechanisms responsible for this effect.

\begin{acknowledgments}

We thank N. Pavloff, M. Isoard and A. Kamchatnov for discussions and T. Macri for discussions and work on nonlocality. This work received funding from the European Union Horizon 2020 research and innovation program under grant agreement No 820392 (PhoQuS) and from the Region Ile-de-France in the framework of DIM SIRTEQ. QG and AB thank the Institut Universitaire de France (IUF) for support. This work was conducted within the framework of the project OPTIMAL granted by the European Union by means of the Fond Européen de développement régional (FEDER).
JG and AP acknowledge financial support from the French ANR under Grant No. ANR-19-CE46-0007 (project ICCI).

\end{acknowledgments}

\appendix

\section{Extraction of the temperature of the gas}

\begin{figure}[h!]
\centerline{{\includegraphics[width=0.4\textwidth]{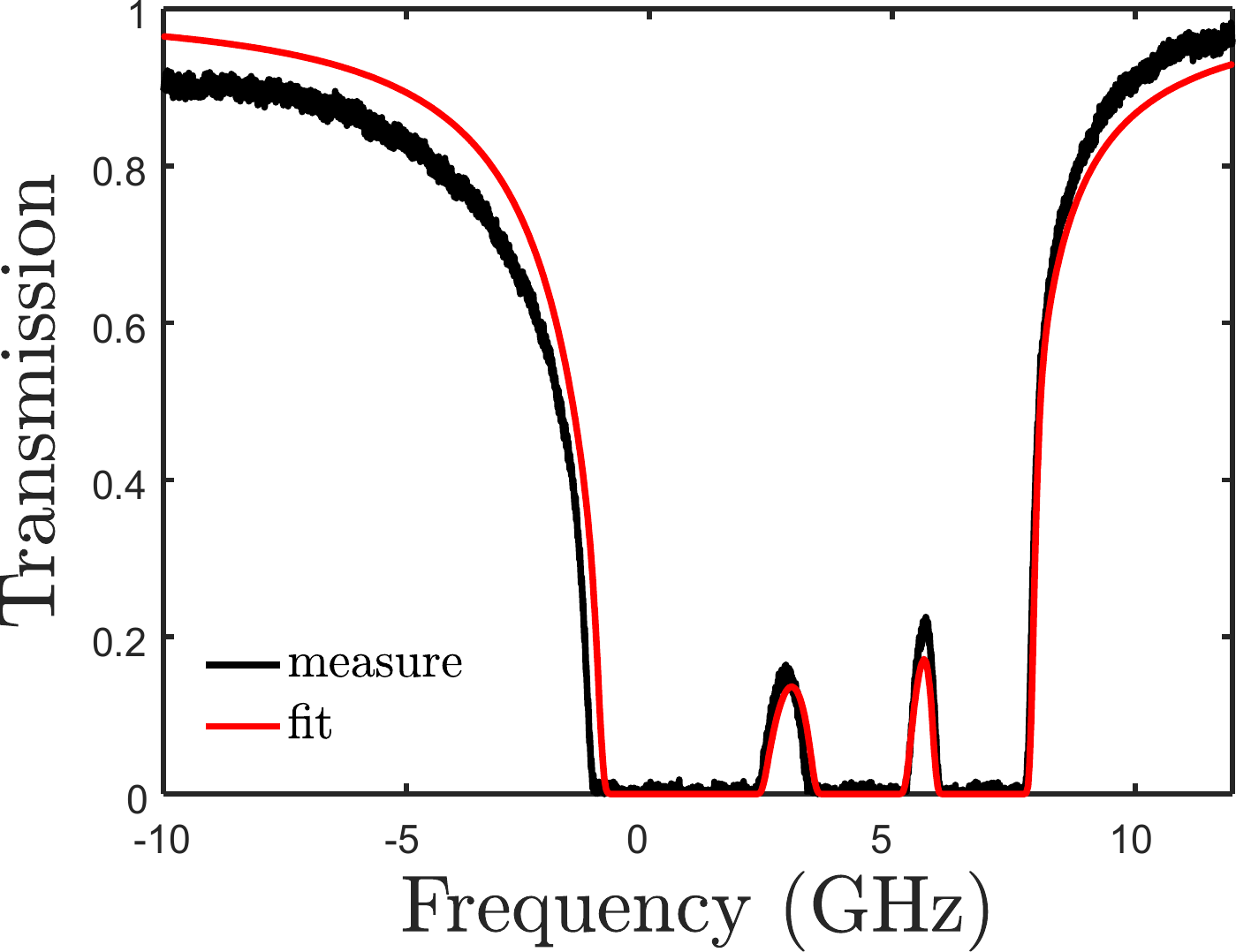}}}
	\caption{Transmission of the laser beam through the cell as a function of the detuning of the laser (same reference as the one of the main text). From the transmission profile a fit to the data allows us to deduce the temperature of the gas. For this particular curve, we find $T = 110 \, ^\circ \text{C}$.}
\label{FigSM1}
\end{figure}

To extract the temperature of the atomic gas, we record the transmission profile of a weak laser beam through the cell as a function of the laser detuning. By fitting this data using a numerical simulation that takes into account the atomic lines of the two isotopes, the Doppler broadening, and the Rubidium vapor pressure as a function of the temperature \cite{siddons2008absolute,agha2011time}, we can infer the temperature of the gas as shown in Fig. \ref{FigSM1}.


\begin{thebibliography}{75}%
\makeatletter
\providecommand \@ifxundefined [1]{%
 \@ifx{#1\undefined}
}%
\providecommand \@ifnum [1]{%
 \ifnum #1\expandafter \@firstoftwo
 \else \expandafter \@secondoftwo
 \fi
}%
\providecommand \@ifx [1]{%
 \ifx #1\expandafter \@firstoftwo
 \else \expandafter \@secondoftwo
 \fi
}%
\providecommand \natexlab [1]{#1}%
\providecommand \enquote  [1]{``#1''}%
\providecommand \bibnamefont  [1]{#1}%
\providecommand \bibfnamefont [1]{#1}%
\providecommand \citenamefont [1]{#1}%
\providecommand \href@noop [0]{\@secondoftwo}%
\providecommand \href [0]{\begingroup \@sanitize@url \@href}%
\providecommand \@href[1]{\@@startlink{#1}\@@href}%
\providecommand \@@href[1]{\endgroup#1\@@endlink}%
\providecommand \@sanitize@url [0]{\catcode `\\12\catcode `\$12\catcode
  `\&12\catcode `\#12\catcode `\^12\catcode `\_12\catcode `\%12\relax}%
\providecommand \@@startlink[1]{}%
\providecommand \@@endlink[0]{}%
\providecommand \url  [0]{\begingroup\@sanitize@url \@url }%
\providecommand \@url [1]{\endgroup\@href {#1}{\urlprefix }}%
\providecommand \urlprefix  [0]{URL }%
\providecommand \Eprint [0]{\href }%
\providecommand \doibase [0]{https://doi.org/}%
\providecommand \selectlanguage [0]{\@gobble}%
\providecommand \bibinfo  [0]{\@secondoftwo}%
\providecommand \bibfield  [0]{\@secondoftwo}%
\providecommand \translation [1]{[#1]}%
\providecommand \BibitemOpen [0]{}%
\providecommand \bibitemStop [0]{}%
\providecommand \bibitemNoStop [0]{.\EOS\space}%
\providecommand \EOS [0]{\spacefactor3000\relax}%
\providecommand \BibitemShut  [1]{\csname bibitem#1\endcsname}%
\let\auto@bib@innerbib\@empty
%</preamble>
\bibitem [{\citenamefont {Landau}\ \emph {et~al.}(1984)\citenamefont {Landau},
  \citenamefont {Lifshitz},\ and\ \citenamefont {Pitaevskii}}]{Landau84}%
  \BibitemOpen
  \bibfield  {author} {\bibinfo {author} {\bibfnamefont {L.~D.}\ \bibnamefont
  {Landau}}, \bibinfo {author} {\bibfnamefont {E.~M.}\ \bibnamefont
  {Lifshitz}},\ and\ \bibinfo {author} {\bibfnamefont {L.~P.}\ \bibnamefont
  {Pitaevskii}},\ }\href@noop {} {\emph {\bibinfo {title} {{Electrodynamics of
  continuous media; 2nd ed.}}}},\ Course of theoretical physics\ (\bibinfo
  {publisher} {Butterworth},\ \bibinfo {address} {Oxford},\ \bibinfo {year}
  {1984})\BibitemShut {NoStop}%
\bibitem [{\citenamefont {Pomeau}\ and\ \citenamefont
  {Rica}(1993)}]{pomeau1993diffraction}%
  \BibitemOpen
  \bibfield  {author} {\bibinfo {author} {\bibfnamefont {Y.}~\bibnamefont
  {Pomeau}}\ and\ \bibinfo {author} {\bibfnamefont {S.}~\bibnamefont {Rica}},\
  }\bibfield  {title} {\bibinfo {title} {Diffraction non lin{\'e}aire},\
  }\href@noop {} {\bibfield  {journal} {\bibinfo  {journal} {Comptes Rendus -
  Académie des Sciences}\ }\textbf {\bibinfo {volume} {317}},\ \bibinfo
  {pages} {1287} (\bibinfo {year} {1993})}\BibitemShut {NoStop}%
\bibitem [{\citenamefont {Frisch}\ \emph {et~al.}(1992)\citenamefont {Frisch},
  \citenamefont {Pomeau},\ and\ \citenamefont {Rica}}]{frisch1992transition}%
  \BibitemOpen
  \bibfield  {author} {\bibinfo {author} {\bibfnamefont {T.}~\bibnamefont
  {Frisch}}, \bibinfo {author} {\bibfnamefont {Y.}~\bibnamefont {Pomeau}},\
  and\ \bibinfo {author} {\bibfnamefont {S.}~\bibnamefont {Rica}},\ }\bibfield
  {title} {\bibinfo {title} {Transition to dissipation in a model of
  superflow},\ }\href@noop {} {\bibfield  {journal} {\bibinfo  {journal}
  {Physical review letters}\ }\textbf {\bibinfo {volume} {69}},\ \bibinfo
  {pages} {1644} (\bibinfo {year} {1992})}\BibitemShut {NoStop}%
\bibitem [{\citenamefont {{Michel}}\ \emph {et~al.}(2018)\citenamefont
  {{Michel}}, \citenamefont {{Boughdad}}, \citenamefont {{Albert}},
  \citenamefont {{Larr{\'e}}},\ and\ \citenamefont {{Bellec}}}]{Michel18}%
  \BibitemOpen
  \bibfield  {author} {\bibinfo {author} {\bibfnamefont {C.}~\bibnamefont
  {{Michel}}}, \bibinfo {author} {\bibfnamefont {O.}~\bibnamefont
  {{Boughdad}}}, \bibinfo {author} {\bibfnamefont {M.}~\bibnamefont
  {{Albert}}}, \bibinfo {author} {\bibfnamefont {P.-{\'E}.}\ \bibnamefont
  {{Larr{\'e}}}},\ and\ \bibinfo {author} {\bibfnamefont {M.}~\bibnamefont
  {{Bellec}}},\ }\bibfield  {title} {\bibinfo {title} {{Superfluid motion and
  drag-force cancellation in a fluid of light}},\ }\href@noop {} {\bibfield
  {journal} {\bibinfo  {journal} {Nature Communications}\ }\textbf {\bibinfo
  {volume} {9}},\ \bibinfo {eid} {2108} (\bibinfo {year} {2018})}\BibitemShut
  {NoStop}%
\bibitem [{\citenamefont {Situ}\ and\ \citenamefont
  {Fleischer}(2020)}]{situ2020dynamics}%
  \BibitemOpen
  \bibfield  {author} {\bibinfo {author} {\bibfnamefont {G.}~\bibnamefont
  {Situ}}\ and\ \bibinfo {author} {\bibfnamefont {J.~W.}\ \bibnamefont
  {Fleischer}},\ }\bibfield  {title} {\bibinfo {title} {Dynamics of the
  {B}erezinskii--{K}osterlitz--{T}houless transition in a photon fluid},\
  }\href@noop {} {\bibfield  {journal} {\bibinfo  {journal} {Nature Photonics}\
  }\textbf {\bibinfo {volume} {14}},\ \bibinfo {pages} {517} (\bibinfo {year}
  {2020})}\BibitemShut {NoStop}%
\bibitem [{\citenamefont {Vocke}\ \emph {et~al.}(2015)\citenamefont {Vocke},
  \citenamefont {Roger}, \citenamefont {Marino}, \citenamefont {Wright},
  \citenamefont {Carusotto}, \citenamefont {Clerici},\ and\ \citenamefont
  {Faccio}}]{Vocke15}%
  \BibitemOpen
  \bibfield  {author} {\bibinfo {author} {\bibfnamefont {D.}~\bibnamefont
  {Vocke}}, \bibinfo {author} {\bibfnamefont {T.}~\bibnamefont {Roger}},
  \bibinfo {author} {\bibfnamefont {F.}~\bibnamefont {Marino}}, \bibinfo
  {author} {\bibfnamefont {E.~M.}\ \bibnamefont {Wright}}, \bibinfo {author}
  {\bibfnamefont {I.}~\bibnamefont {Carusotto}}, \bibinfo {author}
  {\bibfnamefont {M.}~\bibnamefont {Clerici}},\ and\ \bibinfo {author}
  {\bibfnamefont {D.}~\bibnamefont {Faccio}},\ }\bibfield  {title} {\bibinfo
  {title} {Experimental characterization of nonlocal photon fluids},\
  }\href@noop {} {\bibfield  {journal} {\bibinfo  {journal} {Optica}\ }\textbf
  {\bibinfo {volume} {2}},\ \bibinfo {pages} {484} (\bibinfo {year}
  {2015})}\BibitemShut {NoStop}%
\bibitem [{\citenamefont {Vocke}\ \emph {et~al.}(2016)\citenamefont {Vocke},
  \citenamefont {Wilson}, \citenamefont {Marino}, \citenamefont {Carusotto},
  \citenamefont {Wright}, \citenamefont {Roger}, \citenamefont {Anderson},
  \citenamefont {\"Ohberg},\ and\ \citenamefont {Faccio}}]{Vocke16}%
  \BibitemOpen
  \bibfield  {author} {\bibinfo {author} {\bibfnamefont {D.}~\bibnamefont
  {Vocke}}, \bibinfo {author} {\bibfnamefont {K.}~\bibnamefont {Wilson}},
  \bibinfo {author} {\bibfnamefont {F.}~\bibnamefont {Marino}}, \bibinfo
  {author} {\bibfnamefont {I.}~\bibnamefont {Carusotto}}, \bibinfo {author}
  {\bibfnamefont {E.~M.}\ \bibnamefont {Wright}}, \bibinfo {author}
  {\bibfnamefont {T.}~\bibnamefont {Roger}}, \bibinfo {author} {\bibfnamefont
  {B.~P.}\ \bibnamefont {Anderson}}, \bibinfo {author} {\bibfnamefont
  {P.}~\bibnamefont {\"Ohberg}},\ and\ \bibinfo {author} {\bibfnamefont
  {D.}~\bibnamefont {Faccio}},\ }\bibfield  {title} {\bibinfo {title} {Role of
  geometry in the superfluid flow of nonlocal photon fluids},\ }\href@noop {}
  {\bibfield  {journal} {\bibinfo  {journal} {Phys. Rev. A}\ }\textbf {\bibinfo
  {volume} {94}},\ \bibinfo {pages} {013849} (\bibinfo {year}
  {2016})}\BibitemShut {NoStop}%
\bibitem [{\citenamefont {\ifmmode \check{S}\else
  \v{S}\fi{}anti\ifmmode~\acute{c}\else \'{c}\fi{}}\ \emph
  {et~al.}(2018)\citenamefont {\ifmmode \check{S}\else
  \v{S}\fi{}anti\ifmmode~\acute{c}\else \'{c}\fi{}}, \citenamefont {Fusaro},
  \citenamefont {Salem}, \citenamefont {Garnier}, \citenamefont {Picozzi},\
  and\ \citenamefont {Kaiser}}]{Santic18}%
  \BibitemOpen
  \bibfield  {author} {\bibinfo {author} {\bibfnamefont {N.}~\bibnamefont
  {\ifmmode \check{S}\else \v{S}\fi{}anti\ifmmode~\acute{c}\else \'{c}\fi{}}},
  \bibinfo {author} {\bibfnamefont {A.}~\bibnamefont {Fusaro}}, \bibinfo
  {author} {\bibfnamefont {S.}~\bibnamefont {Salem}}, \bibinfo {author}
  {\bibfnamefont {J.}~\bibnamefont {Garnier}}, \bibinfo {author} {\bibfnamefont
  {A.}~\bibnamefont {Picozzi}},\ and\ \bibinfo {author} {\bibfnamefont
  {R.}~\bibnamefont {Kaiser}},\ }\bibfield  {title} {\bibinfo {title}
  {Nonequilibrium precondensation of classical waves in two dimensions
  propagating through atomic vapors},\ }\href@noop {} {\bibfield  {journal}
  {\bibinfo  {journal} {Phys. Rev. Lett.}\ }\textbf {\bibinfo {volume} {120}},\
  \bibinfo {pages} {055301} (\bibinfo {year} {2018})}\BibitemShut {NoStop}%
\bibitem [{\citenamefont {Fontaine}\ \emph {et~al.}(2018)\citenamefont
  {Fontaine}, \citenamefont {Bienaim\'e}, \citenamefont {Pigeon}, \citenamefont
  {Giacobino}, \citenamefont {Bramati},\ and\ \citenamefont
  {Glorieux}}]{Fontaine18}%
  \BibitemOpen
  \bibfield  {author} {\bibinfo {author} {\bibfnamefont {Q.}~\bibnamefont
  {Fontaine}}, \bibinfo {author} {\bibfnamefont {T.}~\bibnamefont
  {Bienaim\'e}}, \bibinfo {author} {\bibfnamefont {S.}~\bibnamefont {Pigeon}},
  \bibinfo {author} {\bibfnamefont {E.}~\bibnamefont {Giacobino}}, \bibinfo
  {author} {\bibfnamefont {A.}~\bibnamefont {Bramati}},\ and\ \bibinfo {author}
  {\bibfnamefont {Q.}~\bibnamefont {Glorieux}},\ }\bibfield  {title} {\bibinfo
  {title} {Observation of the {B}ogoliubov dispersion in a fluid of light},\
  }\href {https://doi.org/10.1103/PhysRevLett.121.183604} {\bibfield  {journal}
  {\bibinfo  {journal} {Phys. Rev. Lett.}\ }\textbf {\bibinfo {volume} {121}},\
  \bibinfo {pages} {183604} (\bibinfo {year} {2018})}\BibitemShut {NoStop}%
\bibitem [{\citenamefont {Fontaine}\ \emph {et~al.}(2020)\citenamefont
  {Fontaine}, \citenamefont {Larr\'e}, \citenamefont {Lerario}, \citenamefont
  {Bienaim\'e}, \citenamefont {Pigeon}, \citenamefont {Faccio}, \citenamefont
  {Carusotto}, \citenamefont {Giacobino}, \citenamefont {Bramati},\ and\
  \citenamefont {Glorieux}}]{Fontaine20}%
  \BibitemOpen
  \bibfield  {author} {\bibinfo {author} {\bibfnamefont {Q.}~\bibnamefont
  {Fontaine}}, \bibinfo {author} {\bibfnamefont {P.-E.}\ \bibnamefont
  {Larr\'e}}, \bibinfo {author} {\bibfnamefont {G.}~\bibnamefont {Lerario}},
  \bibinfo {author} {\bibfnamefont {T.}~\bibnamefont {Bienaim\'e}}, \bibinfo
  {author} {\bibfnamefont {S.}~\bibnamefont {Pigeon}}, \bibinfo {author}
  {\bibfnamefont {D.}~\bibnamefont {Faccio}}, \bibinfo {author} {\bibfnamefont
  {I.}~\bibnamefont {Carusotto}}, \bibinfo {author} {\bibfnamefont
  {E.}~\bibnamefont {Giacobino}}, \bibinfo {author} {\bibfnamefont
  {A.}~\bibnamefont {Bramati}},\ and\ \bibinfo {author} {\bibfnamefont
  {Q.}~\bibnamefont {Glorieux}},\ }\bibfield  {title} {\bibinfo {title}
  {Interferences between {B}ogoliubov excitations in superfluids of light},\
  }\href {https://doi.org/10.1103/PhysRevResearch.2.043297} {\bibfield
  {journal} {\bibinfo  {journal} {Phys. Rev. Research}\ }\textbf {\bibinfo
  {volume} {2}},\ \bibinfo {pages} {043297} (\bibinfo {year}
  {2020})}\BibitemShut {NoStop}%
\bibitem [{\citenamefont {Piekarski}\ \emph {et~al.}(2020)\citenamefont
  {Piekarski}, \citenamefont {Liu}, \citenamefont {Steinhauer}, \citenamefont
  {Giacobino}, \citenamefont {Bramati},\ and\ \citenamefont
  {Glorieux}}]{piekarski2020short}%
  \BibitemOpen
  \bibfield  {author} {\bibinfo {author} {\bibfnamefont {C.}~\bibnamefont
  {Piekarski}}, \bibinfo {author} {\bibfnamefont {W.}~\bibnamefont {Liu}},
  \bibinfo {author} {\bibfnamefont {J.}~\bibnamefont {Steinhauer}}, \bibinfo
  {author} {\bibfnamefont {E.}~\bibnamefont {Giacobino}}, \bibinfo {author}
  {\bibfnamefont {A.}~\bibnamefont {Bramati}},\ and\ \bibinfo {author}
  {\bibfnamefont {Q.}~\bibnamefont {Glorieux}},\ }\bibfield  {title} {\bibinfo
  {title} {Short bragg pulse spectroscopy for a paraxial fluids of light},\
  }\href@noop {} {\bibfield  {journal} {\bibinfo  {journal} {arXiv preprint
  arXiv:2011.12935}\ } (\bibinfo {year} {2020})}\BibitemShut {NoStop}%
\bibitem [{\citenamefont {Whitham}(1974)}]{whitham}%
  \BibitemOpen
  \bibfield  {author} {\bibinfo {author} {\bibfnamefont {G.~B.}\ \bibnamefont
  {Whitham}},\ }\href@noop {} {\emph {\bibinfo {title} {{Linear and Nonlinear
  Waves}}}}\ (\bibinfo  {publisher} {Wiley},\ \bibinfo {year}
  {1974})\BibitemShut {NoStop}%
\bibitem [{\citenamefont {Gurevich}\ and\ \citenamefont
  {Pitaevskii}(1974)}]{gurevich74}%
  \BibitemOpen
  \bibfield  {author} {\bibinfo {author} {\bibfnamefont {A.}~\bibnamefont
  {Gurevich}}\ and\ \bibinfo {author} {\bibfnamefont {L.}~\bibnamefont
  {Pitaevskii}},\ }\bibfield  {title} {\bibinfo {title} {Nonstationary
  structure of a collisionless shock wave},\ }\href@noop {} {\bibfield
  {journal} {\bibinfo  {journal} {Sov. Phys. JETP}\ }\textbf {\bibinfo {volume}
  {38}},\ \bibinfo {pages} {291} (\bibinfo {year} {1974})}\BibitemShut
  {NoStop}%
\bibitem [{\citenamefont {Benjamin}\ and\ \citenamefont
  {Lighthill}(1954)}]{benjamin54}%
  \BibitemOpen
  \bibfield  {author} {\bibinfo {author} {\bibfnamefont {T.~B.}\ \bibnamefont
  {Benjamin}}\ and\ \bibinfo {author} {\bibfnamefont {M.~J.}\ \bibnamefont
  {Lighthill}},\ }\bibfield  {title} {\bibinfo {title} {On cnoidal waves and
  bores},\ }\href@noop {} {\bibfield  {journal} {\bibinfo  {journal} {Proc. R.
  Soc. A}\ }\textbf {\bibinfo {volume} {224}},\ \bibinfo {pages} {448}
  (\bibinfo {year} {1954})}\BibitemShut {NoStop}%
\bibitem [{\citenamefont {Peregrine}(1966)}]{peregrine66}%
  \BibitemOpen
  \bibfield  {author} {\bibinfo {author} {\bibfnamefont {D.~H.}\ \bibnamefont
  {Peregrine}},\ }\bibfield  {title} {\bibinfo {title} {Calculations of the
  development of an undular bore},\ }\href@noop {} {\bibfield  {journal}
  {\bibinfo  {journal} {J. Fluid Mech.}\ }\textbf {\bibinfo {volume} {25}},\
  \bibinfo {pages} {321} (\bibinfo {year} {1966})}\BibitemShut {NoStop}%
\bibitem [{\citenamefont {Johnson}(1972)}]{johnson72}%
  \BibitemOpen
  \bibfield  {author} {\bibinfo {author} {\bibfnamefont {R.~S.}\ \bibnamefont
  {Johnson}},\ }\bibfield  {title} {\bibinfo {title} {Shallow water waves on a
  viscous fluid — the undular bore},\ }\href@noop {} {\bibfield  {journal}
  {\bibinfo  {journal} {Phys. Fluids}\ }\textbf {\bibinfo {volume} {15}},\
  \bibinfo {pages} {1963} (\bibinfo {year} {1972})}\BibitemShut {NoStop}%
\bibitem [{\citenamefont {Sagdeev}(1979)}]{sagdeev79}%
  \BibitemOpen
  \bibfield  {author} {\bibinfo {author} {\bibfnamefont {R.~Z.}\ \bibnamefont
  {Sagdeev}},\ }\bibfield  {title} {\bibinfo {title} {The 1976 oppenheimer
  lectures: Critical problems in plasma astrophysics. ii. singular layers and
  reconnection},\ }\href@noop {} {\bibfield  {journal} {\bibinfo  {journal}
  {Rev. Mod. Phys.}\ }\textbf {\bibinfo {volume} {51}},\ \bibinfo {pages} {11}
  (\bibinfo {year} {1979})}\BibitemShut {NoStop}%
\bibitem [{\citenamefont {Zabusky}\ and\ \citenamefont
  {Kruskal}(1965)}]{zabusky65}%
  \BibitemOpen
  \bibfield  {author} {\bibinfo {author} {\bibfnamefont {N.}~\bibnamefont
  {Zabusky}}\ and\ \bibinfo {author} {\bibfnamefont {M.}~\bibnamefont
  {Kruskal}},\ }\bibfield  {title} {\bibinfo {title} {Interaction of
  ``solitons" in a collisionless plasma and the recurrence of initial states},\
  }\href@noop {} {\bibfield  {journal} {\bibinfo  {journal} {Phys. Rev. Lett.}\
  }\textbf {\bibinfo {volume} {15}},\ \bibinfo {pages} {240} (\bibinfo {year}
  {1965})}\BibitemShut {NoStop}%
\bibitem [{\citenamefont {Taylor}\ \emph {et~al.}(1970)\citenamefont {Taylor},
  \citenamefont {Baker},\ and\ \citenamefont {Ikezi}}]{taylor70}%
  \BibitemOpen
  \bibfield  {author} {\bibinfo {author} {\bibfnamefont {R.}~\bibnamefont
  {Taylor}}, \bibinfo {author} {\bibfnamefont {D.}~\bibnamefont {Baker}},\ and\
  \bibinfo {author} {\bibfnamefont {H.}~\bibnamefont {Ikezi}},\ }\bibfield
  {title} {\bibinfo {title} {Observation of collisionless electrostatic
  shocks},\ }\href@noop {} {\bibfield  {journal} {\bibinfo  {journal} {Phys.
  Rev. Lett.}\ }\textbf {\bibinfo {volume} {24}},\ \bibinfo {pages} {206}
  (\bibinfo {year} {1970})}\BibitemShut {NoStop}%
\bibitem [{\citenamefont {Biondini}\ \emph {et~al.}(2016)\citenamefont
  {Biondini}, \citenamefont {El}, \citenamefont {Hoefer},\ and\ \citenamefont
  {Miller}}]{biondini16}%
  \BibitemOpen
  \bibfield  {author} {\bibinfo {author} {\bibfnamefont {G.}~\bibnamefont
  {Biondini}}, \bibinfo {author} {\bibfnamefont {G.~A.}\ \bibnamefont {El}},
  \bibinfo {author} {\bibfnamefont {M.~A.}\ \bibnamefont {Hoefer}},\ and\
  \bibinfo {author} {\bibfnamefont {P.~D.}\ \bibnamefont {Miller}},\ }\bibfield
   {title} {\bibinfo {title} {Dispersive hydrodynamics},\ }\href@noop {}
  {\bibfield  {journal} {\bibinfo  {journal} {Topical Issue in Physica D}\
  }\textbf {\bibinfo {volume} {333}},\ \bibinfo {pages} {1} (\bibinfo {year}
  {2016})}\BibitemShut {NoStop}%
\bibitem [{\citenamefont {Onorato}\ \emph {et~al.}(2016)\citenamefont
  {Onorato}, \citenamefont {Resitori},\ and\ \citenamefont
  {Baronio}}]{Onorato16}%
  \BibitemOpen
  \bibfield  {author} {\bibinfo {author} {\bibfnamefont {M.}~\bibnamefont
  {Onorato}}, \bibinfo {author} {\bibfnamefont {S.}~\bibnamefont {Resitori}},\
  and\ \bibinfo {author} {\bibfnamefont {F.}~\bibnamefont {Baronio}},\ }\href
  {https://doi.org/10.1007/978-3-319-39214-1} {\emph {\bibinfo {title} {Rogue
  and Shock Waves in Nonlinear Dispersive Media}}},\ Vol.\ \bibinfo {volume}
  {926}\ (\bibinfo {year} {2016})\BibitemShut {NoStop}%
\bibitem [{\citenamefont {El}\ and\ \citenamefont {Hoefer}(2016)}]{el16}%
  \BibitemOpen
  \bibfield  {author} {\bibinfo {author} {\bibfnamefont {G.~A.}\ \bibnamefont
  {El}}\ and\ \bibinfo {author} {\bibfnamefont {M.~A.}\ \bibnamefont
  {Hoefer}},\ }\bibfield  {title} {\bibinfo {title} {Dispersive shock waves and
  modulation theory},\ }\href@noop {} {\bibfield  {journal} {\bibinfo
  {journal} {Physica D}\ }\textbf {\bibinfo {volume} {333}},\ \bibinfo {pages}
  {11} (\bibinfo {year} {2016})}\BibitemShut {NoStop}%
\bibitem [{\citenamefont {Miller}(2016)}]{miller16}%
  \BibitemOpen
  \bibfield  {author} {\bibinfo {author} {\bibfnamefont {P.~D.}\ \bibnamefont
  {Miller}},\ }\bibfield  {title} {\bibinfo {title} {On the generation of
  dispersive shock waves},\ }\href@noop {} {\bibfield  {journal} {\bibinfo
  {journal} {Physica D}\ }\textbf {\bibinfo {volume} {333}},\ \bibinfo {pages}
  {66} (\bibinfo {year} {2016})}\BibitemShut {NoStop}%
\bibitem [{\citenamefont {{Isoard}}\ \emph {et~al.}(2019)\citenamefont
  {{Isoard}}, \citenamefont {{Kamchatnov}},\ and\ \citenamefont
  {{Pavloff}}}]{Isoard19}%
  \BibitemOpen
  \bibfield  {author} {\bibinfo {author} {\bibfnamefont {M.}~\bibnamefont
  {{Isoard}}}, \bibinfo {author} {\bibfnamefont {A.~M.}\ \bibnamefont
  {{Kamchatnov}}},\ and\ \bibinfo {author} {\bibfnamefont {N.}~\bibnamefont
  {{Pavloff}}},\ }\bibfield  {title} {\bibinfo {title} {{Wave breaking and
  formation of dispersive shock waves in a defocusing nonlinear optical
  material}},\ }\href@noop {} {\bibfield  {journal} {\bibinfo  {journal} {arXiv
  e-prints}\ ,\ \bibinfo {eid} {arXiv:1902.06975}} (\bibinfo {year}
  {2019})}\BibitemShut {NoStop}%
\bibitem [{\citenamefont {Dutton}\ \emph {et~al.}(2001)\citenamefont {Dutton},
  \citenamefont {Budde}, \citenamefont {Slowe},\ and\ \citenamefont
  {Hau}}]{Dutton01}%
  \BibitemOpen
  \bibfield  {author} {\bibinfo {author} {\bibfnamefont {Z.}~\bibnamefont
  {Dutton}}, \bibinfo {author} {\bibfnamefont {M.}~\bibnamefont {Budde}},
  \bibinfo {author} {\bibfnamefont {C.}~\bibnamefont {Slowe}},\ and\ \bibinfo
  {author} {\bibfnamefont {L.~V.}\ \bibnamefont {Hau}},\ }\bibfield  {title}
  {\bibinfo {title} {Observation of quantum shock waves created with ultra-
  compressed slow light pulses in a {B}ose-{E}instein condensate},\ }\href
  {https://doi.org/10.1126/science.1062527} {\bibfield  {journal} {\bibinfo
  {journal} {Science}\ }\textbf {\bibinfo {volume} {293}},\ \bibinfo {pages}
  {663} (\bibinfo {year} {2001})}\BibitemShut {NoStop}%
\bibitem [{\citenamefont {Hoefer}\ \emph {et~al.}(2006)\citenamefont {Hoefer},
  \citenamefont {Ablowitz}, \citenamefont {Coddington}, \citenamefont
  {Cornell}, \citenamefont {Engels},\ and\ \citenamefont
  {Schweikhard}}]{hoefer2006dispersive}%
  \BibitemOpen
  \bibfield  {author} {\bibinfo {author} {\bibfnamefont {M.}~\bibnamefont
  {Hoefer}}, \bibinfo {author} {\bibfnamefont {M.}~\bibnamefont {Ablowitz}},
  \bibinfo {author} {\bibfnamefont {I.}~\bibnamefont {Coddington}}, \bibinfo
  {author} {\bibfnamefont {E.~A.}\ \bibnamefont {Cornell}}, \bibinfo {author}
  {\bibfnamefont {P.}~\bibnamefont {Engels}},\ and\ \bibinfo {author}
  {\bibfnamefont {V.}~\bibnamefont {Schweikhard}},\ }\bibfield  {title}
  {\bibinfo {title} {Dispersive and classical shock waves in {B}ose-{E}instein
  condensates and gas dynamics},\ }\href@noop {} {\bibfield  {journal}
  {\bibinfo  {journal} {Physical Review A}\ }\textbf {\bibinfo {volume} {74}},\
  \bibinfo {pages} {023623} (\bibinfo {year} {2006})}\BibitemShut {NoStop}%
\bibitem [{\citenamefont {Chang}\ \emph {et~al.}(2008)\citenamefont {Chang},
  \citenamefont {Engels},\ and\ \citenamefont {Hoefer}}]{chang2008formation}%
  \BibitemOpen
  \bibfield  {author} {\bibinfo {author} {\bibfnamefont {J.~J.}\ \bibnamefont
  {Chang}}, \bibinfo {author} {\bibfnamefont {P.}~\bibnamefont {Engels}},\ and\
  \bibinfo {author} {\bibfnamefont {M.}~\bibnamefont {Hoefer}},\ }\bibfield
  {title} {\bibinfo {title} {Formation of dispersive shock waves by merging and
  splitting {B}ose-{E}instein condensates},\ }\href@noop {} {\bibfield
  {journal} {\bibinfo  {journal} {Physical review letters}\ }\textbf {\bibinfo
  {volume} {101}},\ \bibinfo {pages} {170404} (\bibinfo {year}
  {2008})}\BibitemShut {NoStop}%
\bibitem [{\citenamefont {Trillo}\ \emph {et~al.}(2016)\citenamefont {Trillo},
  \citenamefont {Klein}, \citenamefont {Clauss},\ and\ \citenamefont
  {Onorato}}]{Trillo16}%
  \BibitemOpen
  \bibfield  {author} {\bibinfo {author} {\bibfnamefont {S.}~\bibnamefont
  {Trillo}}, \bibinfo {author} {\bibfnamefont {M.}~\bibnamefont {Klein}},
  \bibinfo {author} {\bibfnamefont {G.}~\bibnamefont {Clauss}},\ and\ \bibinfo
  {author} {\bibfnamefont {M.}~\bibnamefont {Onorato}},\ }\bibfield  {title}
  {\bibinfo {title} {Observation of dispersive shock waves developing from
  initial depressions in shallow water},\ }\href
  {https://doi.org/https://doi.org/10.1016/j.physd.2016.01.007} {\bibfield
  {journal} {\bibinfo  {journal} {Physica D: Nonlinear Phenomena}\ }\textbf
  {\bibinfo {volume} {333}},\ \bibinfo {pages} {276 } (\bibinfo {year}
  {2016})}\BibitemShut {NoStop}%
\bibitem [{\citenamefont {Smyth}\ and\ \citenamefont
  {Holloway}(1988)}]{smyth88}%
  \BibitemOpen
  \bibfield  {author} {\bibinfo {author} {\bibfnamefont {N.~F.}\ \bibnamefont
  {Smyth}}\ and\ \bibinfo {author} {\bibfnamefont {P.~E.}\ \bibnamefont
  {Holloway}},\ }\bibfield  {title} {\bibinfo {title} {Hydraulic jump and
  undular bore formation on a shelf break},\ }\href
  {https://doi.org/10.1175/1520-0485(1988)018<0947:HJAUBF>2.0.CO;2} {\bibfield
  {journal} {\bibinfo  {journal} {Journal of Physical Oceanography}\ }\textbf
  {\bibinfo {volume} {18}},\ \bibinfo {pages} {947 } (\bibinfo {year} {01 Jul.
  1988})}\BibitemShut {NoStop}%
\bibitem [{\citenamefont {Romagnani}\ \emph {et~al.}(2008)\citenamefont
  {Romagnani}, \citenamefont {Bulanov}, \citenamefont {Borghesi}, \citenamefont
  {Audebert}, \citenamefont {Gauthier}, \citenamefont {L\"owenbr\"uck},
  \citenamefont {Mackinnon}, \citenamefont {Patel}, \citenamefont {Pretzler},
  \citenamefont {Toncian},\ and\ \citenamefont {Willi}}]{romagnani08}%
  \BibitemOpen
  \bibfield  {author} {\bibinfo {author} {\bibfnamefont {L.}~\bibnamefont
  {Romagnani}}, \bibinfo {author} {\bibfnamefont {S.~V.}\ \bibnamefont
  {Bulanov}}, \bibinfo {author} {\bibfnamefont {M.}~\bibnamefont {Borghesi}},
  \bibinfo {author} {\bibfnamefont {P.}~\bibnamefont {Audebert}}, \bibinfo
  {author} {\bibfnamefont {J.~C.}\ \bibnamefont {Gauthier}}, \bibinfo {author}
  {\bibfnamefont {K.}~\bibnamefont {L\"owenbr\"uck}}, \bibinfo {author}
  {\bibfnamefont {A.~J.}\ \bibnamefont {Mackinnon}}, \bibinfo {author}
  {\bibfnamefont {P.}~\bibnamefont {Patel}}, \bibinfo {author} {\bibfnamefont
  {G.}~\bibnamefont {Pretzler}}, \bibinfo {author} {\bibfnamefont
  {T.}~\bibnamefont {Toncian}},\ and\ \bibinfo {author} {\bibfnamefont
  {O.}~\bibnamefont {Willi}},\ }\bibfield  {title} {\bibinfo {title}
  {Observation of collisionless shocks in laser-plasma experiments},\ }\href
  {https://doi.org/10.1103/PhysRevLett.101.025004} {\bibfield  {journal}
  {\bibinfo  {journal} {Phys. Rev. Lett.}\ }\textbf {\bibinfo {volume} {101}},\
  \bibinfo {pages} {025004} (\bibinfo {year} {2008})}\BibitemShut {NoStop}%
\bibitem [{\citenamefont {Maiden}\ \emph {et~al.}(2016)\citenamefont {Maiden},
  \citenamefont {Lowman}, \citenamefont {Anderson}, \citenamefont {Schubert},\
  and\ \citenamefont {Hoefer}}]{Maiden16}%
  \BibitemOpen
  \bibfield  {author} {\bibinfo {author} {\bibfnamefont {M.~D.}\ \bibnamefont
  {Maiden}}, \bibinfo {author} {\bibfnamefont {N.~K.}\ \bibnamefont {Lowman}},
  \bibinfo {author} {\bibfnamefont {D.~V.}\ \bibnamefont {Anderson}}, \bibinfo
  {author} {\bibfnamefont {M.~E.}\ \bibnamefont {Schubert}},\ and\ \bibinfo
  {author} {\bibfnamefont {M.~A.}\ \bibnamefont {Hoefer}},\ }\bibfield  {title}
  {\bibinfo {title} {Observation of dispersive shock waves, solitons, and their
  interactions in viscous fluid conduits},\ }\href
  {https://doi.org/10.1103/PhysRevLett.116.174501} {\bibfield  {journal}
  {\bibinfo  {journal} {Phys. Rev. Lett.}\ }\textbf {\bibinfo {volume} {116}},\
  \bibinfo {pages} {174501} (\bibinfo {year} {2016})}\BibitemShut {NoStop}%
\bibitem [{\citenamefont {{Wan}}\ \emph {et~al.}(2007)\citenamefont {{Wan}},
  \citenamefont {{Jia}},\ and\ \citenamefont {{Fleischer}}}]{Wan07}%
  \BibitemOpen
  \bibfield  {author} {\bibinfo {author} {\bibfnamefont {W.}~\bibnamefont
  {{Wan}}}, \bibinfo {author} {\bibfnamefont {S.}~\bibnamefont {{Jia}}},\ and\
  \bibinfo {author} {\bibfnamefont {J.~W.}\ \bibnamefont {{Fleischer}}},\
  }\bibfield  {title} {\bibinfo {title} {{Dispersive superfluid-like shock
  waves in nonlinear optics}},\ }\href {https://doi.org/10.1038/nphys486}
  {\bibfield  {journal} {\bibinfo  {journal} {Nature Physics}\ }\textbf
  {\bibinfo {volume} {3}},\ \bibinfo {pages} {46} (\bibinfo {year}
  {2007})}\BibitemShut {NoStop}%
\bibitem [{\citenamefont {El}\ \emph {et~al.}(2007)\citenamefont {El},
  \citenamefont {Gammal}, \citenamefont {Khamis}, \citenamefont {Kraenkel},\
  and\ \citenamefont {Kamchatnov}}]{Gammal07}%
  \BibitemOpen
  \bibfield  {author} {\bibinfo {author} {\bibfnamefont {G.~A.}\ \bibnamefont
  {El}}, \bibinfo {author} {\bibfnamefont {A.}~\bibnamefont {Gammal}}, \bibinfo
  {author} {\bibfnamefont {E.~G.}\ \bibnamefont {Khamis}}, \bibinfo {author}
  {\bibfnamefont {R.~A.}\ \bibnamefont {Kraenkel}},\ and\ \bibinfo {author}
  {\bibfnamefont {A.~M.}\ \bibnamefont {Kamchatnov}},\ }\bibfield  {title}
  {\bibinfo {title} {Theory of optical dispersive shock waves in
  photorefractive media},\ }\href {https://doi.org/10.1103/PhysRevA.76.053813}
  {\bibfield  {journal} {\bibinfo  {journal} {Phys. Rev. A}\ }\textbf {\bibinfo
  {volume} {76}},\ \bibinfo {pages} {053813} (\bibinfo {year}
  {2007})}\BibitemShut {NoStop}%
\bibitem [{\citenamefont {Ivanov}\ \emph {et~al.}(2020)\citenamefont {Ivanov},
  \citenamefont {Suchorski}, \citenamefont {Kamchatnov}, \citenamefont
  {Isoard},\ and\ \citenamefont {Pavloff}}]{Ivanov20}%
  \BibitemOpen
  \bibfield  {author} {\bibinfo {author} {\bibfnamefont {S.~K.}\ \bibnamefont
  {Ivanov}}, \bibinfo {author} {\bibfnamefont {J.-E.}\ \bibnamefont
  {Suchorski}}, \bibinfo {author} {\bibfnamefont {A.~M.}\ \bibnamefont
  {Kamchatnov}}, \bibinfo {author} {\bibfnamefont {M.}~\bibnamefont {Isoard}},\
  and\ \bibinfo {author} {\bibfnamefont {N.}~\bibnamefont {Pavloff}},\
  }\bibfield  {title} {\bibinfo {title} {Formation of dispersive shock waves in
  a saturable nonlinear medium},\ }\href
  {https://doi.org/10.1103/PhysRevE.102.032215} {\bibfield  {journal} {\bibinfo
   {journal} {Phys. Rev. E}\ }\textbf {\bibinfo {volume} {102}},\ \bibinfo
  {pages} {032215} (\bibinfo {year} {2020})}\BibitemShut {NoStop}%
\bibitem [{\citenamefont {Malaguti}\ \emph {et~al.}(2014)\citenamefont
  {Malaguti}, \citenamefont {Conforti},\ and\ \citenamefont
  {Trillo}}]{Malaguti14}%
  \BibitemOpen
  \bibfield  {author} {\bibinfo {author} {\bibfnamefont {S.}~\bibnamefont
  {Malaguti}}, \bibinfo {author} {\bibfnamefont {M.}~\bibnamefont {Conforti}},\
  and\ \bibinfo {author} {\bibfnamefont {S.}~\bibnamefont {Trillo}},\
  }\bibfield  {title} {\bibinfo {title} {Dispersive radiation induced by shock
  waves in passive resonators},\ }\href {https://doi.org/10.1364/OL.39.005626}
  {\bibfield  {journal} {\bibinfo  {journal} {Opt. Lett.}\ }\textbf {\bibinfo
  {volume} {39}},\ \bibinfo {pages} {5626} (\bibinfo {year}
  {2014})}\BibitemShut {NoStop}%
\bibitem [{\citenamefont {Rothenberg}\ and\ \citenamefont
  {Grischkowsky}(1989)}]{Rothenberg89}%
  \BibitemOpen
  \bibfield  {author} {\bibinfo {author} {\bibfnamefont {J.~E.}\ \bibnamefont
  {Rothenberg}}\ and\ \bibinfo {author} {\bibfnamefont {D.}~\bibnamefont
  {Grischkowsky}},\ }\bibfield  {title} {\bibinfo {title} {Observation of the
  formation of an optical intensity shock and wave breaking in the nonlinear
  propagation of pulses in optical fibers},\ }\href
  {https://doi.org/10.1103/PhysRevLett.62.531} {\bibfield  {journal} {\bibinfo
  {journal} {Phys. Rev. Lett.}\ }\textbf {\bibinfo {volume} {62}},\ \bibinfo
  {pages} {531} (\bibinfo {year} {1989})}\BibitemShut {NoStop}%
\bibitem [{\citenamefont {Xu}\ \emph {et~al.}(2016{\natexlab{a}})\citenamefont
  {Xu}, \citenamefont {Mussot}, \citenamefont {Kudlinski}, \citenamefont
  {Trillo}, \citenamefont {Copie},\ and\ \citenamefont {Conforti}}]{Xu16}%
  \BibitemOpen
  \bibfield  {author} {\bibinfo {author} {\bibfnamefont {G.}~\bibnamefont
  {Xu}}, \bibinfo {author} {\bibfnamefont {A.}~\bibnamefont {Mussot}}, \bibinfo
  {author} {\bibfnamefont {A.}~\bibnamefont {Kudlinski}}, \bibinfo {author}
  {\bibfnamefont {S.}~\bibnamefont {Trillo}}, \bibinfo {author} {\bibfnamefont
  {F.}~\bibnamefont {Copie}},\ and\ \bibinfo {author} {\bibfnamefont
  {M.}~\bibnamefont {Conforti}},\ }\bibfield  {title} {\bibinfo {title} {Shock
  wave generation triggered by a weak background in optical fibers},\ }\href
  {https://doi.org/10.1364/OL.41.002656} {\bibfield  {journal} {\bibinfo
  {journal} {Opt. Lett.}\ }\textbf {\bibinfo {volume} {41}},\ \bibinfo {pages}
  {2656} (\bibinfo {year} {2016}{\natexlab{a}})}\BibitemShut {NoStop}%
\bibitem [{\citenamefont {Fatome}\ \emph {et~al.}(2014)\citenamefont {Fatome},
  \citenamefont {Finot}, \citenamefont {Millot}, \citenamefont {Armaroli},\
  and\ \citenamefont {Trillo}}]{fatome14}%
  \BibitemOpen
  \bibfield  {author} {\bibinfo {author} {\bibfnamefont {J.}~\bibnamefont
  {Fatome}}, \bibinfo {author} {\bibfnamefont {C.}~\bibnamefont {Finot}},
  \bibinfo {author} {\bibfnamefont {G.}~\bibnamefont {Millot}}, \bibinfo
  {author} {\bibfnamefont {A.}~\bibnamefont {Armaroli}},\ and\ \bibinfo
  {author} {\bibfnamefont {S.}~\bibnamefont {Trillo}},\ }\bibfield  {title}
  {\bibinfo {title} {Observation of optical undular bores in multiple four-wave
  mixing},\ }\href {https://doi.org/10.1103/PhysRevX.4.021022} {\bibfield
  {journal} {\bibinfo  {journal} {Phys. Rev. X}\ }\textbf {\bibinfo {volume}
  {4}},\ \bibinfo {pages} {021022} (\bibinfo {year} {2014})}\BibitemShut
  {NoStop}%
\bibitem [{\citenamefont {Garnier}\ \emph {et~al.}(2013)\citenamefont
  {Garnier}, \citenamefont {Xu}, \citenamefont {Trillo},\ and\ \citenamefont
  {Picozzi}}]{garnier13}%
  \BibitemOpen
  \bibfield  {author} {\bibinfo {author} {\bibfnamefont {J.}~\bibnamefont
  {Garnier}}, \bibinfo {author} {\bibfnamefont {G.}~\bibnamefont {Xu}},
  \bibinfo {author} {\bibfnamefont {S.}~\bibnamefont {Trillo}},\ and\ \bibinfo
  {author} {\bibfnamefont {A.}~\bibnamefont {Picozzi}},\ }\bibfield  {title}
  {\bibinfo {title} {Incoherent dispersive shocks in the spectral evolution of
  random waves},\ }\href {https://doi.org/10.1103/PhysRevLett.111.113902}
  {\bibfield  {journal} {\bibinfo  {journal} {Phys. Rev. Lett.}\ }\textbf
  {\bibinfo {volume} {111}},\ \bibinfo {pages} {113902} (\bibinfo {year}
  {2013})}\BibitemShut {NoStop}%
\bibitem [{\citenamefont {Wetzel}\ \emph {et~al.}(2016)\citenamefont {Wetzel},
  \citenamefont {Bongiovanni}, \citenamefont {Kues}, \citenamefont {Hu},
  \citenamefont {Chen}, \citenamefont {Trillo}, \citenamefont {Dudley},
  \citenamefont {Wabnitz},\ and\ \citenamefont {Morandotti}}]{wetzel16}%
  \BibitemOpen
  \bibfield  {author} {\bibinfo {author} {\bibfnamefont {B.}~\bibnamefont
  {Wetzel}}, \bibinfo {author} {\bibfnamefont {D.}~\bibnamefont {Bongiovanni}},
  \bibinfo {author} {\bibfnamefont {M.}~\bibnamefont {Kues}}, \bibinfo {author}
  {\bibfnamefont {Y.}~\bibnamefont {Hu}}, \bibinfo {author} {\bibfnamefont
  {Z.}~\bibnamefont {Chen}}, \bibinfo {author} {\bibfnamefont {S.}~\bibnamefont
  {Trillo}}, \bibinfo {author} {\bibfnamefont {J.~M.}\ \bibnamefont {Dudley}},
  \bibinfo {author} {\bibfnamefont {S.}~\bibnamefont {Wabnitz}},\ and\ \bibinfo
  {author} {\bibfnamefont {R.}~\bibnamefont {Morandotti}},\ }\bibfield  {title}
  {\bibinfo {title} {Experimental generation of riemann waves in optics: A
  route to shock wave control},\ }\href
  {https://doi.org/10.1103/PhysRevLett.117.073902} {\bibfield  {journal}
  {\bibinfo  {journal} {Phys. Rev. Lett.}\ }\textbf {\bibinfo {volume} {117}},\
  \bibinfo {pages} {073902} (\bibinfo {year} {2016})}\BibitemShut {NoStop}%
\bibitem [{\citenamefont {Trillo}\ and\ \citenamefont
  {Conforti}(2017)}]{Trillo17}%
  \BibitemOpen
  \bibfield  {author} {\bibinfo {author} {\bibfnamefont {S.}~\bibnamefont
  {Trillo}}\ and\ \bibinfo {author} {\bibfnamefont {M.}~\bibnamefont
  {Conforti}},\ }\href {https://doi.org/10.1007/978-981-10-1477-2_16-1} {\emph
  {\bibinfo {title} {Handbook of Optical Fibers}}},\ edited by\ \bibinfo
  {editor} {\bibfnamefont {G.-D.}\ \bibnamefont {Peng}}\ (\bibinfo  {publisher}
  {Springer Singapore},\ \bibinfo {address} {Singapore},\ \bibinfo {year}
  {2017})\ pp.\ \bibinfo {pages} {1--48}\BibitemShut {NoStop}%
\bibitem [{\citenamefont {Xu}\ \emph {et~al.}(2017)\citenamefont {Xu},
  \citenamefont {Conforti}, \citenamefont {Kudlinski}, \citenamefont {Mussot},\
  and\ \citenamefont {Trillo}}]{Xu17}%
  \BibitemOpen
  \bibfield  {author} {\bibinfo {author} {\bibfnamefont {G.}~\bibnamefont
  {Xu}}, \bibinfo {author} {\bibfnamefont {M.}~\bibnamefont {Conforti}},
  \bibinfo {author} {\bibfnamefont {A.}~\bibnamefont {Kudlinski}}, \bibinfo
  {author} {\bibfnamefont {A.}~\bibnamefont {Mussot}},\ and\ \bibinfo {author}
  {\bibfnamefont {S.}~\bibnamefont {Trillo}},\ }\bibfield  {title} {\bibinfo
  {title} {Dispersive dam-break flow of a photon fluid},\ }\href
  {https://doi.org/10.1103/PhysRevLett.118.254101} {\bibfield  {journal}
  {\bibinfo  {journal} {Phys. Rev. Lett.}\ }\textbf {\bibinfo {volume} {118}},\
  \bibinfo {pages} {254101} (\bibinfo {year} {2017})}\BibitemShut {NoStop}%
\bibitem [{\citenamefont {Ghofraniha}\ \emph {et~al.}(2007)\citenamefont
  {Ghofraniha}, \citenamefont {Conti}, \citenamefont {Ruocco},\ and\
  \citenamefont {Trillo}}]{Ghofraniha07}%
  \BibitemOpen
  \bibfield  {author} {\bibinfo {author} {\bibfnamefont {N.}~\bibnamefont
  {Ghofraniha}}, \bibinfo {author} {\bibfnamefont {C.}~\bibnamefont {Conti}},
  \bibinfo {author} {\bibfnamefont {G.}~\bibnamefont {Ruocco}},\ and\ \bibinfo
  {author} {\bibfnamefont {S.}~\bibnamefont {Trillo}},\ }\bibfield  {title}
  {\bibinfo {title} {Shocks in nonlocal media},\ }\href
  {https://doi.org/10.1103/PhysRevLett.99.043903} {\bibfield  {journal}
  {\bibinfo  {journal} {Phys. Rev. Lett.}\ }\textbf {\bibinfo {volume} {99}},\
  \bibinfo {pages} {043903} (\bibinfo {year} {2007})}\BibitemShut {NoStop}%
\bibitem [{\citenamefont {Conti}\ \emph {et~al.}(2009)\citenamefont {Conti},
  \citenamefont {Fratalocchi}, \citenamefont {Peccianti}, \citenamefont
  {Ruocco},\ and\ \citenamefont {Trillo}}]{Conti09}%
  \BibitemOpen
  \bibfield  {author} {\bibinfo {author} {\bibfnamefont {C.}~\bibnamefont
  {Conti}}, \bibinfo {author} {\bibfnamefont {A.}~\bibnamefont {Fratalocchi}},
  \bibinfo {author} {\bibfnamefont {M.}~\bibnamefont {Peccianti}}, \bibinfo
  {author} {\bibfnamefont {G.}~\bibnamefont {Ruocco}},\ and\ \bibinfo {author}
  {\bibfnamefont {S.}~\bibnamefont {Trillo}},\ }\bibfield  {title} {\bibinfo
  {title} {Observation of a gradient catastrophe generating solitons},\ }\href
  {https://doi.org/10.1103/PhysRevLett.102.083902} {\bibfield  {journal}
  {\bibinfo  {journal} {Phys. Rev. Lett.}\ }\textbf {\bibinfo {volume} {102}},\
  \bibinfo {pages} {083902} (\bibinfo {year} {2009})}\BibitemShut {NoStop}%
\bibitem [{\citenamefont {Ghofraniha}\ \emph {et~al.}(2012)\citenamefont
  {Ghofraniha}, \citenamefont {Amato}, \citenamefont {Folli}, \citenamefont
  {Trillo}, \citenamefont {DelRe},\ and\ \citenamefont {Conti}}]{Ghofraniha12}%
  \BibitemOpen
  \bibfield  {author} {\bibinfo {author} {\bibfnamefont {N.}~\bibnamefont
  {Ghofraniha}}, \bibinfo {author} {\bibfnamefont {L.~S.}\ \bibnamefont
  {Amato}}, \bibinfo {author} {\bibfnamefont {V.}~\bibnamefont {Folli}},
  \bibinfo {author} {\bibfnamefont {S.}~\bibnamefont {Trillo}}, \bibinfo
  {author} {\bibfnamefont {E.}~\bibnamefont {DelRe}},\ and\ \bibinfo {author}
  {\bibfnamefont {C.}~\bibnamefont {Conti}},\ }\bibfield  {title} {\bibinfo
  {title} {Measurement of scaling laws for shock waves in thermal nonlocal
  media},\ }\href {https://doi.org/10.1364/OL.37.002325} {\bibfield  {journal}
  {\bibinfo  {journal} {Opt. Lett.}\ }\textbf {\bibinfo {volume} {37}},\
  \bibinfo {pages} {2325} (\bibinfo {year} {2012})}\BibitemShut {NoStop}%
\bibitem [{\citenamefont {Gentilini}\ \emph {et~al.}(2015)\citenamefont
  {Gentilini}, \citenamefont {Braidotti}, \citenamefont {Marcucci},
  \citenamefont {DelRe},\ and\ \citenamefont {Conti}}]{gentilini15}%
  \BibitemOpen
  \bibfield  {author} {\bibinfo {author} {\bibfnamefont {S.}~\bibnamefont
  {Gentilini}}, \bibinfo {author} {\bibfnamefont {M.~C.}\ \bibnamefont
  {Braidotti}}, \bibinfo {author} {\bibfnamefont {G.}~\bibnamefont {Marcucci}},
  \bibinfo {author} {\bibfnamefont {E.}~\bibnamefont {DelRe}},\ and\ \bibinfo
  {author} {\bibfnamefont {C.}~\bibnamefont {Conti}},\ }\bibfield  {title}
  {\bibinfo {title} {Nonlinear gamow vectors, shock waves, and irreversibility
  in optically nonlocal media},\ }\href
  {https://doi.org/10.1103/PhysRevA.92.023801} {\bibfield  {journal} {\bibinfo
  {journal} {Phys. Rev. A}\ }\textbf {\bibinfo {volume} {92}},\ \bibinfo
  {pages} {023801} (\bibinfo {year} {2015})}\BibitemShut {NoStop}%
\bibitem [{\citenamefont {Braidotti}\ \emph {et~al.}(2016)\citenamefont
  {Braidotti}, \citenamefont {Gentilini},\ and\ \citenamefont
  {Conti}}]{braidotti16}%
  \BibitemOpen
  \bibfield  {author} {\bibinfo {author} {\bibfnamefont {M.~C.}\ \bibnamefont
  {Braidotti}}, \bibinfo {author} {\bibfnamefont {S.}~\bibnamefont
  {Gentilini}},\ and\ \bibinfo {author} {\bibfnamefont {C.}~\bibnamefont
  {Conti}},\ }\bibfield  {title} {\bibinfo {title} {Gamow vectors explain the
  shock profile},\ }\href {https://doi.org/10.1364/OE.24.021963} {\bibfield
  {journal} {\bibinfo  {journal} {Opt. Express}\ }\textbf {\bibinfo {volume}
  {24}},\ \bibinfo {pages} {21963} (\bibinfo {year} {2016})}\BibitemShut
  {NoStop}%
\bibitem [{\citenamefont {El}\ and\ \citenamefont {Smyth}(2016)}]{el16b}%
  \BibitemOpen
  \bibfield  {author} {\bibinfo {author} {\bibfnamefont {G.~A.}\ \bibnamefont
  {El}}\ and\ \bibinfo {author} {\bibfnamefont {N.~F.}\ \bibnamefont {Smyth}},\
  }\bibfield  {title} {\bibinfo {title} {Radiating dispersive shock waves in
  non-local optical media},\ }\href@noop {} {\bibfield  {journal} {\bibinfo
  {journal} {Proc. R. Soc.A}\ }\textbf {\bibinfo {volume} {472}},\ \bibinfo
  {pages} {20150633} (\bibinfo {year} {2016})}\BibitemShut {NoStop}%
\bibitem [{\citenamefont {Karpov}\ \emph {et~al.}(2015)\citenamefont {Karpov},
  \citenamefont {Congy}, \citenamefont {Sivan}, \citenamefont {Fleurov},
  \citenamefont {Pavloff},\ and\ \citenamefont {Bar-Ad}}]{Karpov15}%
  \BibitemOpen
  \bibfield  {author} {\bibinfo {author} {\bibfnamefont {M.}~\bibnamefont
  {Karpov}}, \bibinfo {author} {\bibfnamefont {T.}~\bibnamefont {Congy}},
  \bibinfo {author} {\bibfnamefont {Y.}~\bibnamefont {Sivan}}, \bibinfo
  {author} {\bibfnamefont {V.}~\bibnamefont {Fleurov}}, \bibinfo {author}
  {\bibfnamefont {N.}~\bibnamefont {Pavloff}},\ and\ \bibinfo {author}
  {\bibfnamefont {S.}~\bibnamefont {Bar-Ad}},\ }\bibfield  {title} {\bibinfo
  {title} {Spontaneously formed autofocusing caustics in a confined
  self-defocusing medium},\ }\href {https://doi.org/10.1364/OPTICA.2.001053}
  {\bibfield  {journal} {\bibinfo  {journal} {Optica}\ }\textbf {\bibinfo
  {volume} {2}},\ \bibinfo {pages} {1053} (\bibinfo {year} {2015})}\BibitemShut
  {NoStop}%
\bibitem [{\citenamefont {Marcucci}\ \emph {et~al.}(2019)\citenamefont
  {Marcucci}, \citenamefont {Pierangeli}, \citenamefont {Gentilini},
  \citenamefont {Ghofraniha}, \citenamefont {Chen}, ,\ and\ \citenamefont
  {Conti}}]{marcucci19}%
  \BibitemOpen
  \bibfield  {author} {\bibinfo {author} {\bibfnamefont {G.}~\bibnamefont
  {Marcucci}}, \bibinfo {author} {\bibfnamefont {D.}~\bibnamefont
  {Pierangeli}}, \bibinfo {author} {\bibfnamefont {S.}~\bibnamefont
  {Gentilini}}, \bibinfo {author} {\bibfnamefont {N.}~\bibnamefont
  {Ghofraniha}}, \bibinfo {author} {\bibfnamefont {Z.}~\bibnamefont {Chen}}, ,\
  and\ \bibinfo {author} {\bibfnamefont {C.}~\bibnamefont {Conti}},\ }\bibfield
   {title} {\bibinfo {title} {Optical spatial shock waves in nonlocal nonlinear
  media},\ }\href@noop {} {\bibfield  {journal} {\bibinfo  {journal} {Adv.
  Phys. X}\ }\textbf {\bibinfo {volume} {4}},\ \bibinfo {pages} {1662733}
  (\bibinfo {year} {2019})}\BibitemShut {NoStop}%
\bibitem [{\citenamefont {Marcucci}\ \emph {et~al.}(2020)\citenamefont
  {Marcucci}, \citenamefont {Hu}, \citenamefont {Cala}, \citenamefont {Man},
  \citenamefont {Pierangeli}, \citenamefont {Conti},\ and\ \citenamefont
  {Chen}}]{marcucci20}%
  \BibitemOpen
  \bibfield  {author} {\bibinfo {author} {\bibfnamefont {G.}~\bibnamefont
  {Marcucci}}, \bibinfo {author} {\bibfnamefont {X.}~\bibnamefont {Hu}},
  \bibinfo {author} {\bibfnamefont {P.}~\bibnamefont {Cala}}, \bibinfo {author}
  {\bibfnamefont {W.}~\bibnamefont {Man}}, \bibinfo {author} {\bibfnamefont
  {D.}~\bibnamefont {Pierangeli}}, \bibinfo {author} {\bibfnamefont
  {C.}~\bibnamefont {Conti}},\ and\ \bibinfo {author} {\bibfnamefont
  {Z.}~\bibnamefont {Chen}},\ }\bibfield  {title} {\bibinfo {title}
  {Anisotropic optical shock waves in isotropic media with giant nonlocal
  nonlinearity},\ }\href {https://doi.org/10.1103/PhysRevLett.125.243902}
  {\bibfield  {journal} {\bibinfo  {journal} {Phys. Rev. Lett.}\ }\textbf
  {\bibinfo {volume} {125}},\ \bibinfo {pages} {243902} (\bibinfo {year}
  {2020})}\BibitemShut {NoStop}%
\bibitem [{\citenamefont {Xu}\ \emph {et~al.}(2015)\citenamefont {Xu},
  \citenamefont {Vocke}, \citenamefont {Faccio}, \citenamefont {Garnier},
  \citenamefont {Roger}, \citenamefont {Trillo},\ and\ \citenamefont
  {Picozzi}}]{Xu15}%
  \BibitemOpen
  \bibfield  {author} {\bibinfo {author} {\bibfnamefont {G.}~\bibnamefont
  {Xu}}, \bibinfo {author} {\bibfnamefont {D.}~\bibnamefont {Vocke}}, \bibinfo
  {author} {\bibfnamefont {D.}~\bibnamefont {Faccio}}, \bibinfo {author}
  {\bibfnamefont {J.}~\bibnamefont {Garnier}}, \bibinfo {author} {\bibfnamefont
  {T.}~\bibnamefont {Roger}}, \bibinfo {author} {\bibfnamefont
  {S.}~\bibnamefont {Trillo}},\ and\ \bibinfo {author} {\bibfnamefont
  {A.}~\bibnamefont {Picozzi}},\ }\bibfield  {title} {\bibinfo {title} {From
  coherent shocklets to giant collective incoherent shock waves in nonlocal
  turbulent flows},\ }\href@noop {} {\bibfield  {journal} {\bibinfo  {journal}
  {Nature communications}\ }\textbf {\bibinfo {volume} {6}},\ \bibinfo {pages}
  {8131} (\bibinfo {year} {2015})}\BibitemShut {NoStop}%
\bibitem [{\citenamefont {Skupin}\ \emph {et~al.}(2007)\citenamefont {Skupin},
  \citenamefont {Saffman},\ and\ \citenamefont
  {Krolikowski}}]{skupin2007nonlocal}%
  \BibitemOpen
  \bibfield  {author} {\bibinfo {author} {\bibfnamefont {S.}~\bibnamefont
  {Skupin}}, \bibinfo {author} {\bibfnamefont {M.}~\bibnamefont {Saffman}},\
  and\ \bibinfo {author} {\bibfnamefont {W.}~\bibnamefont {Krolikowski}},\
  }\bibfield  {title} {\bibinfo {title} {Nonlocal stabilization of nonlinear
  beams in a self-focusing atomic vapor},\ }\href@noop {} {\bibfield  {journal}
  {\bibinfo  {journal} {Physical review letters}\ }\textbf {\bibinfo {volume}
  {98}},\ \bibinfo {pages} {263902} (\bibinfo {year} {2007})}\BibitemShut
  {NoStop}%
\bibitem [{\citenamefont {Tanemura}\ \emph {et~al.}(2004)\citenamefont
  {Tanemura}, \citenamefont {Ozeki},\ and\ \citenamefont
  {Kikuchi}}]{tanemura04}%
  \BibitemOpen
  \bibfield  {author} {\bibinfo {author} {\bibfnamefont {T.}~\bibnamefont
  {Tanemura}}, \bibinfo {author} {\bibfnamefont {Y.}~\bibnamefont {Ozeki}},\
  and\ \bibinfo {author} {\bibfnamefont {K.}~\bibnamefont {Kikuchi}},\
  }\bibfield  {title} {\bibinfo {title} {Modulational instability and
  parametric amplification induced by loss dispersion in optical fibers},\
  }\href {https://doi.org/10.1103/PhysRevLett.93.163902} {\bibfield  {journal}
  {\bibinfo  {journal} {Phys. Rev. Lett.}\ }\textbf {\bibinfo {volume} {93}},\
  \bibinfo {pages} {163902} (\bibinfo {year} {2004})}\BibitemShut {NoStop}%
\bibitem [{\citenamefont {Perego}\ \emph {et~al.}(2018)\citenamefont {Perego},
  \citenamefont {Turitsyn},\ and\ \citenamefont {Staliunas}}]{perego18}%
  \BibitemOpen
  \bibfield  {author} {\bibinfo {author} {\bibfnamefont {A.~M.}\ \bibnamefont
  {Perego}}, \bibinfo {author} {\bibfnamefont {S.~K.}\ \bibnamefont
  {Turitsyn}},\ and\ \bibinfo {author} {\bibfnamefont {K.}~\bibnamefont
  {Staliunas}},\ }\bibfield  {title} {\bibinfo {title} {Gain through losses in
  nonlinear optics},\ }\href@noop {} {\bibfield  {journal} {\bibinfo  {journal}
  {Light: Science \& Applications}\ }\textbf {\bibinfo {volume} {7}},\ \bibinfo
  {pages} {43} (\bibinfo {year} {2018})}\BibitemShut {NoStop}%
\bibitem [{\citenamefont {Tikhonenko}\ \emph {et~al.}(1996)\citenamefont
  {Tikhonenko}, \citenamefont {Christou}, \citenamefont {Luther-Davies},\ and\
  \citenamefont {Kivshar}}]{Tikhonenko96}%
  \BibitemOpen
  \bibfield  {author} {\bibinfo {author} {\bibfnamefont {V.}~\bibnamefont
  {Tikhonenko}}, \bibinfo {author} {\bibfnamefont {J.}~\bibnamefont
  {Christou}}, \bibinfo {author} {\bibfnamefont {B.}~\bibnamefont
  {Luther-Davies}},\ and\ \bibinfo {author} {\bibfnamefont {Y.~S.}\
  \bibnamefont {Kivshar}},\ }\bibfield  {title} {\bibinfo {title} {Observation
  of vortex solitons created by the instability of dark soliton stripes},\
  }\href {https://doi.org/10.1364/OL.21.001129} {\bibfield  {journal} {\bibinfo
   {journal} {Opt. Lett.}\ }\textbf {\bibinfo {volume} {21}},\ \bibinfo {pages}
  {1129} (\bibinfo {year} {1996})}\BibitemShut {NoStop}%
\bibitem [{\citenamefont {Swartzlander}\ and\ \citenamefont
  {Law}(1992)}]{Swartzlander92}%
  \BibitemOpen
  \bibfield  {author} {\bibinfo {author} {\bibfnamefont {G.~A.}\ \bibnamefont
  {Swartzlander}}\ and\ \bibinfo {author} {\bibfnamefont {C.~T.}\ \bibnamefont
  {Law}},\ }\bibfield  {title} {\bibinfo {title} {Optical vortex solitons
  observed in kerr nonlinear media},\ }\href
  {https://doi.org/10.1103/PhysRevLett.69.2503} {\bibfield  {journal} {\bibinfo
   {journal} {Phys. Rev. Lett.}\ }\textbf {\bibinfo {volume} {69}},\ \bibinfo
  {pages} {2503} (\bibinfo {year} {1992})}\BibitemShut {NoStop}%
\bibitem [{\citenamefont {Abuzarli}\ \emph {et~al.}(2021)\citenamefont
  {Abuzarli}, \citenamefont {Bienaim{\'e}}, \citenamefont {Giacobino},
  \citenamefont {Bramati},\ and\ \citenamefont {Glorieux}}]{abuzarli2021blast}%
  \BibitemOpen
  \bibfield  {author} {\bibinfo {author} {\bibfnamefont {M.}~\bibnamefont
  {Abuzarli}}, \bibinfo {author} {\bibfnamefont {T.}~\bibnamefont
  {Bienaim{\'e}}}, \bibinfo {author} {\bibfnamefont {E.}~\bibnamefont
  {Giacobino}}, \bibinfo {author} {\bibfnamefont {A.}~\bibnamefont {Bramati}},\
  and\ \bibinfo {author} {\bibfnamefont {Q.}~\bibnamefont {Glorieux}},\
  }\bibfield  {title} {\bibinfo {title} {Blast waves in a paraxial fluid of
  light},\ }\href@noop {} {\bibfield  {journal} {\bibinfo  {journal} {arXiv
  preprint arXiv:2101.09040}\ } (\bibinfo {year} {2021})}\BibitemShut {NoStop}%
\bibitem [{\citenamefont {Bienaim\'e}\ \emph {et~al.}(2021)\citenamefont
  {Bienaim\'e}, \citenamefont {Isoard}, \citenamefont {Fontaine}, \citenamefont
  {Bramati}, \citenamefont {Kamchatnov}, \citenamefont {Glorieux},\ and\
  \citenamefont {Pavloff}}]{bienaime2021controlled}%
  \BibitemOpen
  \bibfield  {author} {\bibinfo {author} {\bibfnamefont {T.}~\bibnamefont
  {Bienaim\'e}}, \bibinfo {author} {\bibfnamefont {M.}~\bibnamefont {Isoard}},
  \bibinfo {author} {\bibfnamefont {Q.}~\bibnamefont {Fontaine}}, \bibinfo
  {author} {\bibfnamefont {A.}~\bibnamefont {Bramati}}, \bibinfo {author}
  {\bibfnamefont {A.~M.}\ \bibnamefont {Kamchatnov}}, \bibinfo {author}
  {\bibfnamefont {Q.}~\bibnamefont {Glorieux}},\ and\ \bibinfo {author}
  {\bibfnamefont {N.}~\bibnamefont {Pavloff}},\ }\bibfield  {title} {\bibinfo
  {title} {Quantitative analysis of shock wave dynamics in a fluid of light},\
  }\href {https://doi.org/10.1103/PhysRevLett.126.183901} {\bibfield  {journal}
  {\bibinfo  {journal} {Phys. Rev. Lett.}\ }\textbf {\bibinfo {volume} {126}},\
  \bibinfo {pages} {183901} (\bibinfo {year} {2021})}\BibitemShut {NoStop}%
\bibitem [{\citenamefont {Baranov}(2008)}]{baranov08}%
  \BibitemOpen
  \bibfield  {author} {\bibinfo {author} {\bibfnamefont {M.}~\bibnamefont
  {Baranov}},\ }\bibfield  {title} {\bibinfo {title} {Theoretical progress in
  many-body physics with ultracold dipolar gases},\ }\href
  {https://doi.org/https://doi.org/10.1016/j.physrep.2008.04.007} {\bibfield
  {journal} {\bibinfo  {journal} {Physics Reports}\ }\textbf {\bibinfo {volume}
  {464}},\ \bibinfo {pages} {71} (\bibinfo {year} {2008})}\BibitemShut
  {NoStop}%
\bibitem [{\citenamefont {Peccianti}\ and\ \citenamefont
  {Assanto}(2012)}]{peccianti12}%
  \BibitemOpen
  \bibfield  {author} {\bibinfo {author} {\bibfnamefont {M.}~\bibnamefont
  {Peccianti}}\ and\ \bibinfo {author} {\bibfnamefont {G.}~\bibnamefont
  {Assanto}},\ }\bibfield  {title} {\bibinfo {title} {Nematicons},\ }\href
  {https://doi.org/https://doi.org/10.1016/j.physrep.2012.02.004} {\bibfield
  {journal} {\bibinfo  {journal} {Physics Reports}\ }\textbf {\bibinfo {volume}
  {516}},\ \bibinfo {pages} {147} (\bibinfo {year} {2012})}\BibitemShut
  {NoStop}%
\bibitem [{\citenamefont {Rotschild}\ \emph {et~al.}(2006)\citenamefont
  {Rotschild}, \citenamefont {Alfassi}, \citenamefont {Cohen},\ and\
  \citenamefont {Segev}}]{Rotschild06}%
  \BibitemOpen
  \bibfield  {author} {\bibinfo {author} {\bibfnamefont {C.}~\bibnamefont
  {Rotschild}}, \bibinfo {author} {\bibfnamefont {B.}~\bibnamefont {Alfassi}},
  \bibinfo {author} {\bibfnamefont {O.}~\bibnamefont {Cohen}},\ and\ \bibinfo
  {author} {\bibfnamefont {M.}~\bibnamefont {Segev}},\ }\bibfield  {title}
  {\bibinfo {title} {Long-range interactions between optical solitons},\
  }\href@noop {} {\bibfield  {journal} {\bibinfo  {journal} {Nature Physics}\
  }\textbf {\bibinfo {volume} {2}},\ \bibinfo {pages} {769} (\bibinfo {year}
  {2006})}\BibitemShut {NoStop}%
\bibitem [{\citenamefont {Zakharov}\ \emph {et~al.}(1985)\citenamefont
  {Zakharov}, \citenamefont {Musher},\ and\ \citenamefont
  {Rubenchik}}]{zakharov85}%
  \BibitemOpen
  \bibfield  {author} {\bibinfo {author} {\bibfnamefont {V.}~\bibnamefont
  {Zakharov}}, \bibinfo {author} {\bibfnamefont {S.}~\bibnamefont {Musher}},\
  and\ \bibinfo {author} {\bibfnamefont {A.}~\bibnamefont {Rubenchik}},\
  }\bibfield  {title} {\bibinfo {title} {Hamiltonian approach to the
  description of non-linear plasma phenomena},\ }\href
  {https://doi.org/https://doi.org/10.1016/0370-1573(85)90040-7} {\bibfield
  {journal} {\bibinfo  {journal} {Physics Reports}\ }\textbf {\bibinfo {volume}
  {129}},\ \bibinfo {pages} {285} (\bibinfo {year} {1985})}\BibitemShut
  {NoStop}%
\bibitem [{\citenamefont {Zhang}\ \emph {et~al.}(2015)\citenamefont {Zhang},
  \citenamefont {Cheng}, \citenamefont {Yin}, \citenamefont {Bai},
  \citenamefont {Zhao},\ and\ \citenamefont {Ren}}]{Zhang15}%
  \BibitemOpen
  \bibfield  {author} {\bibinfo {author} {\bibfnamefont {Y.}~\bibnamefont
  {Zhang}}, \bibinfo {author} {\bibfnamefont {X.}~\bibnamefont {Cheng}},
  \bibinfo {author} {\bibfnamefont {X.}~\bibnamefont {Yin}}, \bibinfo {author}
  {\bibfnamefont {J.}~\bibnamefont {Bai}}, \bibinfo {author} {\bibfnamefont
  {P.}~\bibnamefont {Zhao}},\ and\ \bibinfo {author} {\bibfnamefont
  {Z.}~\bibnamefont {Ren}},\ }\bibfield  {title} {\bibinfo {title} {Research of
  far-field diffraction intensity pattern in hot atomic rb sample},\ }\href
  {https://doi.org/10.1364/OE.23.005468} {\bibfield  {journal} {\bibinfo
  {journal} {Opt. Express}\ }\textbf {\bibinfo {volume} {23}},\ \bibinfo
  {pages} {5468} (\bibinfo {year} {2015})}\BibitemShut {NoStop}%
\bibitem [{sup()}]{supplemental}%
  \BibitemOpen
  \href@noop {} {\bibinfo {title} {See supplementary material}}\BibitemShut
  {NoStop}%
\bibitem [{\citenamefont {Evans}(2002)}]{evans}%
  \BibitemOpen
  \bibfield  {author} {\bibinfo {author} {\bibfnamefont {L.~C.}\ \bibnamefont
  {Evans}},\ }\href@noop {} {\emph {\bibinfo {title} {{Partial Differential
  Equations; 2nd ed.}}}}\ (\bibinfo  {publisher} {AMS},\ \bibinfo {address}
  {Providence},\ \bibinfo {year} {2002})\BibitemShut {NoStop}%
\bibitem [{\citenamefont {Xu}\ \emph {et~al.}(2016{\natexlab{b}})\citenamefont
  {Xu}, \citenamefont {Garnier}, \citenamefont {Faccio}, \citenamefont
  {Trillo},\ and\ \citenamefont {Picozzi}}]{xu16incoherent}%
  \BibitemOpen
  \bibfield  {author} {\bibinfo {author} {\bibfnamefont {G.}~\bibnamefont
  {Xu}}, \bibinfo {author} {\bibfnamefont {J.}~\bibnamefont {Garnier}},
  \bibinfo {author} {\bibfnamefont {D.}~\bibnamefont {Faccio}}, \bibinfo
  {author} {\bibfnamefont {S.}~\bibnamefont {Trillo}},\ and\ \bibinfo {author}
  {\bibfnamefont {A.}~\bibnamefont {Picozzi}},\ }\bibfield  {title} {\bibinfo
  {title} {Incoherent shock waves in long-range optical turbulence},\ }\href
  {https://doi.org/https://doi.org/10.1016/j.physd.2016.02.015} {\bibfield
  {journal} {\bibinfo  {journal} {Physica D: Nonlinear Phenomena}\ }\textbf
  {\bibinfo {volume} {333}},\ \bibinfo {pages} {310} (\bibinfo {year}
  {2016}{\natexlab{b}})},\ \bibinfo {note} {dispersive
  Hydrodynamics}\BibitemShut {NoStop}%
\bibitem [{\citenamefont {Xu}\ \emph {et~al.}(2018)\citenamefont {Xu},
  \citenamefont {Fusaro}, \citenamefont {Garnier},\ and\ \citenamefont
  {Picozzi}}]{xu18incoherent}%
  \BibitemOpen
  \bibfield  {author} {\bibinfo {author} {\bibfnamefont {G.}~\bibnamefont
  {Xu}}, \bibinfo {author} {\bibfnamefont {A.}~\bibnamefont {Fusaro}}, \bibinfo
  {author} {\bibfnamefont {J.}~\bibnamefont {Garnier}},\ and\ \bibinfo {author}
  {\bibfnamefont {A.}~\bibnamefont {Picozzi}},\ }\bibfield  {title} {\bibinfo
  {title} {Incoherent shock and collapse singularities in non-instantaneous
  nonlinear media},\ }\bibfield  {journal} {\bibinfo  {journal} {Applied
  Sciences}\ }\textbf {\bibinfo {volume} {8}},\ \href
  {https://doi.org/10.3390/app8122559} {10.3390/app8122559} (\bibinfo {year}
  {2018})\BibitemShut {NoStop}%
\bibitem [{\citenamefont {Fusaro}\ \emph {et~al.}(2017)\citenamefont {Fusaro},
  \citenamefont {Garnier}, \citenamefont {Xu}, \citenamefont {Conti},
  \citenamefont {Faccio}, \citenamefont {Trillo},\ and\ \citenamefont
  {Picozzi}}]{fusaro17}%
  \BibitemOpen
  \bibfield  {author} {\bibinfo {author} {\bibfnamefont {A.}~\bibnamefont
  {Fusaro}}, \bibinfo {author} {\bibfnamefont {J.}~\bibnamefont {Garnier}},
  \bibinfo {author} {\bibfnamefont {G.}~\bibnamefont {Xu}}, \bibinfo {author}
  {\bibfnamefont {C.}~\bibnamefont {Conti}}, \bibinfo {author} {\bibfnamefont
  {D.}~\bibnamefont {Faccio}}, \bibinfo {author} {\bibfnamefont
  {S.}~\bibnamefont {Trillo}},\ and\ \bibinfo {author} {\bibfnamefont
  {A.}~\bibnamefont {Picozzi}},\ }\bibfield  {title} {\bibinfo {title}
  {Emergence of long-range phase coherence in nonlocal fluids of light},\
  }\href {https://doi.org/10.1103/PhysRevA.95.063818} {\bibfield  {journal}
  {\bibinfo  {journal} {Phys. Rev. A}\ }\textbf {\bibinfo {volume} {95}},\
  \bibinfo {pages} {063818} (\bibinfo {year} {2017})}\BibitemShut {NoStop}%
\bibitem [{\citenamefont {Tam}\ and\ \citenamefont
  {Happer}(1977)}]{tam1977long}%
  \BibitemOpen
  \bibfield  {author} {\bibinfo {author} {\bibfnamefont {A.}~\bibnamefont
  {Tam}}\ and\ \bibinfo {author} {\bibfnamefont {W.}~\bibnamefont {Happer}},\
  }\bibfield  {title} {\bibinfo {title} {Long-range interactions between cw
  self-focused laser beams in an atomic vapor},\ }\href@noop {} {\bibfield
  {journal} {\bibinfo  {journal} {Physical Review Letters}\ }\textbf {\bibinfo
  {volume} {38}},\ \bibinfo {pages} {278} (\bibinfo {year} {1977})}\BibitemShut
  {NoStop}%
\bibitem [{\citenamefont {Suter}\ and\ \citenamefont
  {Blasberg}(1993)}]{suter1993stabilization}%
  \BibitemOpen
  \bibfield  {author} {\bibinfo {author} {\bibfnamefont {D.}~\bibnamefont
  {Suter}}\ and\ \bibinfo {author} {\bibfnamefont {T.}~\bibnamefont
  {Blasberg}},\ }\bibfield  {title} {\bibinfo {title} {Stabilization of
  transverse solitary waves by a nonlocal response of the nonlinear medium},\
  }\href@noop {} {\bibfield  {journal} {\bibinfo  {journal} {Physical Review
  A}\ }\textbf {\bibinfo {volume} {48}},\ \bibinfo {pages} {4583} (\bibinfo
  {year} {1993})}\BibitemShut {NoStop}%
\bibitem [{\citenamefont {Maucher}\ \emph {et~al.}(2016)\citenamefont
  {Maucher}, \citenamefont {Pohl}, \citenamefont {Skupin},\ and\ \citenamefont
  {Krolikowski}}]{Maucher_2016}%
  \BibitemOpen
  \bibfield  {author} {\bibinfo {author} {\bibfnamefont {F.}~\bibnamefont
  {Maucher}}, \bibinfo {author} {\bibfnamefont {T.}~\bibnamefont {Pohl}},
  \bibinfo {author} {\bibfnamefont {S.}~\bibnamefont {Skupin}},\ and\ \bibinfo
  {author} {\bibfnamefont {W.}~\bibnamefont {Krolikowski}},\ }\bibfield
  {title} {\bibinfo {title} {Self-organization of light in optical media with
  competing nonlinearities},\ }\bibfield  {journal} {\bibinfo  {journal}
  {Physical Review Letters}\ }\textbf {\bibinfo {volume} {116}},\ \href
  {https://doi.org/10.1103/physrevlett.116.163902}
  {10.1103/physrevlett.116.163902} (\bibinfo {year} {2016})\BibitemShut
  {NoStop}%
\bibitem [{\citenamefont {van Kampen}\ \emph {et~al.}(1997)\citenamefont {van
  Kampen}, \citenamefont {Sautenkov}, \citenamefont {Shalagin}, \citenamefont
  {Eliel},\ and\ \citenamefont {Woerdman}}]{VanKampen97}%
  \BibitemOpen
  \bibfield  {author} {\bibinfo {author} {\bibfnamefont {H.}~\bibnamefont {van
  Kampen}}, \bibinfo {author} {\bibfnamefont {V.~A.}\ \bibnamefont
  {Sautenkov}}, \bibinfo {author} {\bibfnamefont {A.~M.}\ \bibnamefont
  {Shalagin}}, \bibinfo {author} {\bibfnamefont {E.~R.}\ \bibnamefont
  {Eliel}},\ and\ \bibinfo {author} {\bibfnamefont {J.~P.}\ \bibnamefont
  {Woerdman}},\ }\bibfield  {title} {\bibinfo {title} {Dipole-dipole
  collision-induced transport of resonance excitation in a high-density atomic
  vapor},\ }\href {https://doi.org/10.1103/PhysRevA.56.3569} {\bibfield
  {journal} {\bibinfo  {journal} {Phys. Rev. A}\ }\textbf {\bibinfo {volume}
  {56}},\ \bibinfo {pages} {3569} (\bibinfo {year} {1997})}\BibitemShut
  {NoStop}%
\bibitem [{\citenamefont {Siddons}\ \emph {et~al.}(2008)\citenamefont
  {Siddons}, \citenamefont {Adams}, \citenamefont {Ge},\ and\ \citenamefont
  {Hughes}}]{siddons2008absolute}%
  \BibitemOpen
  \bibfield  {author} {\bibinfo {author} {\bibfnamefont {P.}~\bibnamefont
  {Siddons}}, \bibinfo {author} {\bibfnamefont {C.~S.}\ \bibnamefont {Adams}},
  \bibinfo {author} {\bibfnamefont {C.}~\bibnamefont {Ge}},\ and\ \bibinfo
  {author} {\bibfnamefont {I.~G.}\ \bibnamefont {Hughes}},\ }\bibfield  {title}
  {\bibinfo {title} {Absolute absorption on rubidium d lines: comparison
  between theory and experiment},\ }\href@noop {} {\bibfield  {journal}
  {\bibinfo  {journal} {Journal of Physics B: Atomic, Molecular and Optical
  Physics}\ }\textbf {\bibinfo {volume} {41}},\ \bibinfo {pages} {155004}
  (\bibinfo {year} {2008})}\BibitemShut {NoStop}%
\bibitem [{\citenamefont {Agha}\ \emph {et~al.}(2011)\citenamefont {Agha},
  \citenamefont {Giarmatzi}, \citenamefont {Glorieux}, \citenamefont
  {Coudreau}, \citenamefont {Grangier},\ and\ \citenamefont
  {Messin}}]{agha2011time}%
  \BibitemOpen
  \bibfield  {author} {\bibinfo {author} {\bibfnamefont {I.~H.}\ \bibnamefont
  {Agha}}, \bibinfo {author} {\bibfnamefont {C.}~\bibnamefont {Giarmatzi}},
  \bibinfo {author} {\bibfnamefont {Q.}~\bibnamefont {Glorieux}}, \bibinfo
  {author} {\bibfnamefont {T.}~\bibnamefont {Coudreau}}, \bibinfo {author}
  {\bibfnamefont {P.}~\bibnamefont {Grangier}},\ and\ \bibinfo {author}
  {\bibfnamefont {G.}~\bibnamefont {Messin}},\ }\bibfield  {title} {\bibinfo
  {title} {Time-resolved detection of relative-intensity squeezed nanosecond
  pulses in an 87rb vapor},\ }\href@noop {} {\bibfield  {journal} {\bibinfo
  {journal} {New Journal of Physics}\ }\textbf {\bibinfo {volume} {13}},\
  \bibinfo {pages} {043030} (\bibinfo {year} {2011})}\BibitemShut {NoStop}%
\end{thebibliography}
\end{document}